\documentstyle [12pt,epsfig]{article} 
\makeatletter
\@addtoreset{equation}{section}
\makeatother

\newcommand{\ba}{\begin{eqnarray}}
\newcommand{\ea}{\end{eqnarray}}

\begin{document}
\newcommand{\BS}{\bigskip}
\newcommand{\SECTION}[1]{\BS{\large\section{\bf #1}}}
\newcommand{\SUBSECTION}[1]{\BS{\large\subsection{\bf #1}}}
\newcommand{\SUBSUBSECTION}[1]{\BS{\large\subsubsection{\bf #1}}}

\newcommand {\rbf}  {\overline{r}_f}
\newcommand {\vbf}  {\overline{v}_f}
\newcommand {\abf}  {\overline{a}_f}
\newcommand {\sbf}  {\overline{s}_f}
\newcommand {\sbq}  {\overline{s}_q}
\newcommand {\vbq}  {\overline{v}_q}
\newcommand {\abq}  {\overline{a}_q}
\newcommand {\rbQ}  {\overline{r}_Q}
\newcommand {\vbQ}  {\overline{v}_Q}
\newcommand {\abQ}  {\overline{a}_Q}
\newcommand {\rbl}  {\overline{r}_l}

\newcommand {\vbl}  {\overline{v}_l}
\newcommand {\abl}  {\overline{a}_l}
\newcommand {\sbl}  {\overline{s}_l}
\newcommand {\rbc}  {\overline{r}_c}
\newcommand {\vbc}  {\overline{v}_c}
\newcommand {\abc}  {\overline{a}_c}
\newcommand {\sbc}  {\overline{s}_c}
\newcommand {\rbb}  {\overline{r}_b}
\newcommand {\vbb}  {\overline{v}_b}
\newcommand {\abb}  {\overline{a}_b}
\newcommand {\sbb}  {\overline{s}_b}
\newcommand {\gbl}  {\overline{g}_b^L}
\newcommand {\gbr}  {\overline{g}_b^R}
\newcommand {\gcl}  {\overline{g}_c^L}
\newcommand {\gcr}  {\overline{g}_c^R}
\newcommand {\Afbf}  {A_{FB}^{0,f}}
\newcommand {\Afbl}  {A_{FB}^{0,l}}
\newcommand {\Afbc}  {A_{FB}^{0,c}}
\newcommand {\Afbb}  {A_{FB}^{0,b}}
\newcommand {\tpol}  {$\tau$-polarisation}
\newcommand {\alps}  {\alpha_s(M_Z)}
\begin{titlepage}
\hspace*{8cm} {UGVA-DPNC 1998/07-178 July 1998}
\newline
 {\it Revised and corrected version of UGVA-DPNC 1997/10-173, CERN OPEN 97-037,
 hep-ph/9801403. To be published in Physical Review D.}
%\hspace*{8cm}
\begin{center}
\vspace*{2cm}
{\large \bf
A MODEL INDEPENDENT ANALYSIS OF LEP AND SLD 
DATA ON Z-DECAYS; IS THE STANDARD MODEL CONFIRMED?}
\vspace*{1.5cm}
\end{center}
\begin{center}
{\bf J.H.Field }
\end{center}
\begin{center}
{ 
D\'{e}partement de Physique Nucl\'{e}aire et Corpusculaire
 Universit\'{e} de Gen\`{e}ve . 24, quai Ernest-Ansermet
 CH-1211 Gen\`{e}ve 4.
}
\end{center}
\vspace*{2cm}
\begin{abstract}
 A model independent analysis has been performed on the
 LEP and SLD data on Z decays. Using only very weak 
 theoretical assumptions, the effective vector and 
 axial-vector couplings of leptons, c quarks and b quarks 
 have been extracted. Although the lepton and c quark couplings
 agree well with Standard Model predictions, those of the b quark
 show deviations of more than three standard deviations. The effect
 is mainly in the right-handed b quark coupling, the left-handed
 coupling being consistent (at the 2$\sigma$ level)
  with the Standard Model prediction. 
 The probability that the observed 
 deviations of all the measured effective couplings are statistical
 fluctuations from lepton universality and the Standard Model
 is estimated to be 0.9$\%$. The estimated probability that
 the deviations in the leptonic and
  b quark couplings alone are a fluctuation is 0.18$\%$.
 A thorough discussion 
is made of the internal consistency of the different measurements
 contributing to the average values $A_l$ and $A_b$ used to 
  extract the b quark couplings, as well as possible sources of
 systematic error that may not, hitherto, have been taken into
  account. Excluding \tpol~ measurements, which show internal 
 inconsistencies, from the averages increases the deviations of the
  extracted b quark couplings from the Standard Model predictions to the 
 four standard deviation level.  
               
\end{abstract}
\vspace*{1cm}
\newline 
{\it Keywords ;} Standard Electroweak Model, LEP and SLD data, Z-decays,
 Anomalous right-handed b quark coupling.
\end{titlepage}
\SECTION{\bf{Introduction}}
This study is based on a recent compilation~\cite{x1} of experimental
results on Z decays. The aim is to answer the question: `are the data 
consistent with the predictions of the Standard Electroweak Model (SM)
~\cite{x2}' To this end the analysis is carried out in three independent steps.
In the first, the data is used to extract the effective vector ($\overline{v}$)
and axial vector ($\overline{a}$) coupling constants
of the charged leptons (assuming lepton universality) and of 
the c and b quarks. This is done using only weak theoretical assumptions.
The effective couplings are then compared with the predictions of the
SM and the confidence levels (CLs) for consistency of the measurements with the SM
are calculated. In the second step the experimental data contributing to
 deviations observed from the SM in the b quark couplings are 
 critically examined. Issues
 addressed include the r\^{o}les of statistical and systematic errors, as well
 as the internal consistency of physical parameters measured using different
 experimental methods. Finally the observed deviations from the SM are 
 assumed to represent real physical effects, whose interpretation is discussed.
 The three steps described above constitute
 the material of the following three Sections. A summary and  
 outlook are given in the final Section.
 
\SECTION{\bf{Extraction of the Effective Weak Coupling Constants}}
 It is convenient to define the following auxiliary quantities\footnote{
 The fermion masses are set to zero in Eqn.(2.2). Only for the b quark
 do the fermion mass terms give a non-negligible contribution}:
 \begin{eqnarray}
 \rbf &\equiv &\vbf/\abf \\
 \sbf &\equiv &(\abf)^2+(\vbf)^2
 \end{eqnarray}
 that may be simply derived from the measurements. The experimental errors on 
 $\rbf$ and $\sbf$ ($f$ here stands for lepton or quark) are,
 unlike those in $\vbf$ and $\abf$, essentially
  uncorrelated, simplifying the calculation of the statistical significance
  of any deviations observed from the SM expectations.
  Throughout this Section, the SM predictions quoted are those of 
  the global SM fit with $m_t = 172$ GeV, $m_H = 149$ GeV reported in Ref.[1].
  The effect of varying the Higgs boson mass, the only remaining unknown
  parameter of the SM, is discussed in Section 4 below. 
  \par The quantities $\rbf$ ($f = l,c,b$)\footnote{Unless otherwise stated,
  $l$ is a generic lepton label and $e-\mu-\tau$ universality is assumed}
  and $\sbl$ may be directly obtained from the data without any additional
  assumptions concerning the poorly-measured~\cite{x3} couplings of the u, d, s
  quarks. A further assumption is, however, necessary in order to extract 
  $\sbc$, $\sbb$ and hence the c and b quark couplings. In order to perform
  an analysis which is, as far as possible, `model independent', and to avoid
  the specific assumption of the validity of the SM, as used in the fits of
  Ref.[1], the weaker hypothesis of quark-lepton universality is made for
  the fermions e, $\mu$, $\tau$, u, d, s, c. That is, all these fermions
  are assumed to have the same effective weak mixing angle~\cite{x1}.
  The derived values of $\vbc$ and $\abc$ presented below are found to be,
  within errors, in good agreement with this hypothesis. Another possibility
  is to assume a value of $\alps$ derived from non electroweak-related 
  measurements. In this case the c and b quark couplings may be derived
  from the ratios $\Gamma_Q/\Gamma_l$ ( Q $=$ c,b) without any assumption
  concerning the couplings of the light quarks. There is now, however, the
  disadvantage that the extracted values of the electroweak couplings are
   strongly correlated with the assumed $\alps$ value.
  \par At LEP, $\rbf$ is found from the measured, corrected, pole
  forward/backward charge 
  asymmetries $\Afbf$~\cite{x1} via the relations:
  \begin{eqnarray}
  \Afbf & = & \frac{3}{4} A_e A_f     \\
  A_f &\equiv & \frac{2 \vbf \abf}{(\abf)^2+(\vbf)^2}= \frac{2 \rbf}{1+\rbf^2}
  \end{eqnarray}
  $A_e$ and $A_{\tau}$ have also been measured at LEP via the angular
   dependence of the $\tau$-polarisation asymmetry:
   \begin{equation}
   \overline{P}_{\tau}(\cos \theta) = -\frac{A_{\tau}+A_e F(\theta)}
   {1+A_{\tau} A_e F(\theta)}
   \end{equation}
   where 
   \[ F(\theta) \equiv 2 \cos \theta /(1+ \cos^2 \theta) \]
   and $\theta$ is the angle between the incoming $e^-$ and the outgoing 
   $\tau^-$ in the $\tau$-pair centre of mass frame.
   At SLD, $A_e$ is directly measured by the left/right beam polarisation 
   asymmetry $A_{LR}$, while $A_c$ and $A_b$ are determined from the
   left/right-forward/backward asymmetries of tagged heavy quarks.
   \par The separate LEP and SLD average values of the 
   electroweak observables, which are 
   directly sensitive to the effective couplings, are reported in 
   Table 1. The combined LEP/SLD averages of $A_l$, $A_c$, $A_b$, $R_c$
   and $R_b$, where $R_Q= \Gamma_Q/\Gamma_{had}$ , ( $Q =$ c, b) are reported
   in Table 2. 
   \par It may be remarked that,
    while there is good agreement between the different values of
    $A_l$ derived from $\Afbl$ ($l = e, \mu, \tau$) using
    Eqn.(2.3) and that derived from $A_{LR}$,
    ( weighted average 0.1533(27)\footnote{Throughout this paper
    total experimental errors are given in terms of the last
    significant figures. 0.1533(27) denotes 0.1533$\pm$ 0.0027},
      $\chi^2 = 3.85$ for 3 DOF , CL =  28$\%$), the values of both 
      $A_e$ and $A_{\tau}$ derived from the $\tau$-polarisation measurements
    are significantly lower. In fact, the average value of $A_l$ from
    $\tau$-polarisation lies 2.5 $\sigma$ below the average from $\Afbl$,
    $A_{LR}$. Including, or not including, the $\tau$-polarisation data
    changes the LEP/SLD average value of $A_l$ by more than one standard
    deviation (see the first column of Table 2).
    Because of this possible inconsistency in the measured $A_l$ values
    from different sources,
    the extraction of the coupling constants will be done
    throughout this paper using values
    of $A_l$ that either include, or exclude, the $\tau$ polarisation
    results. Significant differences are found. Possible explanations
    for the apparent inconsistencies in the $A_l$ measurements are
    discussed in the following Section. 
    \par The values of $\rbf$ ($f = l, c, b$) derived from the measured
    values of $A_l$ using Eqn.(2.4) are presented in Table 3.
     For the b quark, mass effects
     were taken into account by using the corrected form of Eqn.(2.4):
    \begin{equation}
    A_b = \frac{2 (\sqrt{1-4 \mu_b}) \rbb}{1-4 \mu_b+(1+2 \mu_b) \rbb^2}
    \end{equation} 
    where $\mu_b = (\overline{m}_b(M_Z)/M_Z)^2 \simeq 1.0 \times 10^{-3}$. The
    running b quark mass is taken as $\overline{m}_b(M_Z) = 3.0$ GeV~\cite{x4}.
    Agreement is seen with the SM at the $2 \sigma$ level for $\rbl$, at
     $< 1 \sigma$ for $\rbc$, but only at the $3.3 \sigma$ level for $\rbb$.
     The similar discrepancy for $A_b$ was mentioned, but not discussed
     in terms of the b quark couplings, in Ref.[1]. If the \tpol~measurements 
     of $A_l$ are excluded from the average, the discrepancy of $\rbl$ with the
     SM approaches 3$\sigma$\footnote{As shown in Section 4 below, the
      discrepancies with the SM predictions for the leptonic couplings,
      unlike those of the b quark, are reduced by assuming a smaller 
      value of $m_H$ and a larger value of $m_t$
     than that found in the global fit of Ref.[1]}
      and that of $\rbb$ exceeds 4$\sigma$.
     \par The quantity $\sbl$ is derived from the leptonic width $\Gamma_l$
     using the relation:
     \begin{equation}
     \sbl = (\abl)^2+(\vbl)^2 = \frac{12 \pi \Gamma_l}{\sqrt{2} G_{\mu} M_Z^3}
     \frac{1}{(1+\frac{3 \alpha(M_Z)}{4 \pi})}
 \end{equation}
 The value obtained for $\sbl$, quoted in the first column
  of Table 4, uses the LEP average value of $\Gamma_l$ from Table 1 together with:
 $G_{\mu} = 1.16639 \times 10^{-5}$ (GeV)$^2$~\cite{x5}, $M_Z = 91.1863$ GeV, 
 and $\alpha(M_Z)^{-1} = 128.896$~\cite{x1}. Good agreement is found with the
 SM value. Solving Eqns.(2.1) and (2.2) for $\abl$ and $\vbl$ yields the results
 presented in Table 5. As in the calculation of all
 the other effective couplings, the signs of $\abl$ and $\vbl$ are chosen to be
 the same as the SM predictions. The values of $\abl$ and $\vbl$ are in
 good agreement with the LEP+SLD averages quoted in Ref.[1], taking into account
 the slightly different analysis procedures\footnote{Ref.[1] included small
 mass corrections in calculating $\abl$ and $\vbl$ which are neglected here.}.
 Both $\abl$ and $\vbl$ are in agreement with the SM predictions.
 \par The quantities $\overline{s}_Q$ ( $Q =$ c,b), including quark mass
  effects, may be derived from the measured quantities $R_Q$ via the
  relation:  
  \begin{equation}
 \overline{s}_Q = (\overline{a}_Q)^2(1-6 \mu_Q)+(\overline{v}_Q)^2=
 \frac{R_Q S_Q}{(1-R_Q)C_Q^{QED}C_Q^{QCD}} 
 \end{equation}
 where
 \[ S_Q \equiv \sum_{q \ne Q}[(\overline{a}_q)^2(1-6 \mu_q)+(\overline{v}_q)^2]\]
 and~\cite{x6}:
 \begin{eqnarray}
 C_Q^i & = & 1+\delta_Q^i - < \delta_{q \ne Q}^i >~~~~~(i = QED, QCD),~~~~ \mu_q
 = 0~~{\rm for}~~ q \ne b \nonumber \\
  \delta_q^{QED} & = & \frac{3 (e_q)^2}{4 \pi} \alpha (M_Z),~~
  \delta_{q \ne b}^{QCD} = 1.00a_s+1.42a_s^2,~~~
  \delta_b^{QCD} = .99a_s-1.55a_s^2  \nonumber
  \end{eqnarray}
  $q$ is a generic quark flavour index, $e_q$ the quark electric charge in
  units of that of the positron
  and $a_s \equiv \alps/\pi$ .
  $<X>$ denotes the quark flavour average of $X$.
    As mentioned above, $\mu_b = 1.0 \times 10^{-3}$
  while, taking into account the present experimental error on $R_c$, $\mu_c$
  is set to zero. The numerical values of the QED and QCD correction factors,
  with $\alpha_s(M_Z) = 0.12$ and $\alpha(M_Z)^{-1} = 128.9$, are presented in 
  Table 6. The non-b quark couplings in Eqn.(2.8) are written, conventionally,
   as:
  \begin{eqnarray}
   \overline{a}_q & = & \sqrt{\rho_q} ~T^q_3  \\
   \overline{v}_q & = & \sqrt{\rho_q} (T^q_3-2 e_q (\overline{s}^q_W)^2)
   \end{eqnarray} 
   where, assuming non-b quark lepton universality
   \footnote{ Here the weak isospin symmetry of the SM is invoked to
   calculate the unobserved couplings. It is also assumed that the quantum
   corrections contained in $\rho_q$ and $(\overline{s}^q_W)^2$, though
   not necessarily those of the SM, are universal.}:
  \begin{eqnarray}
   \sqrt{\rho_q} & = & \sqrt{\rho_l}~ =  ~2 |\abl |~~~(~{\rm all}~ q \ne b~)  \\
   (\overline{s}^q_W)^2 & = & \frac{1}{4}(1-\rbl)~~~~~~~~(~{\rm all}~ q \ne b~)
 \end{eqnarray}
 and $T_3^q$ is the third component of the weak isospin of the quark q.
 Substituting the measured values of $\rbl$, $\abl$, from Tables 3, 5 and
 of $R_c$, $R_b$ from Table 2, leads to the values of $\sbc$, $\sbb$ reported
  in Table 4. Note that the value of $\sbb$, and hence $\abb$ and $\vbb$ are
  extracted first. The latter are then substituted into Eqn.(2.8) (taking into
  account their experimental errors) in order to find $\sbc$.
In Table 4 good agreement is seen between the measured values of 
$\sbl$ and $\sbc$ and the SM predictions. On the other hand, $\sbb$ lies 
1.3 $\sigma$ above the prediction, a residual of 
the well known `$R_b$ problem'~\cite{x1}.  
   Solving Eqns.(2.1) and (2.8) then gives the effective coupling
  constants for the heavy quarks
  presented in Table 7. The values 
  found, as well as the errors, agree well with those reported by Renton
  in a recent review~\cite{x7}.
The solutions for $\abf$, $\vbf$ obtained from the essentially uncorrelated
quantities $\rbf$ and $\sbf$ are shown graphically in Figs1a,1b,1c for 
 f $=$ $l$,c,b respectively. The corresponding solutions when the \tpol~ 
measurements of $A_l$ are excluded from the average are shown in Figs.2a,2b,2c.
It is clear from Figs1c,2c that largest discrepancy with the SM is in the
parameter $\rbb$(completely determined by $A_b$) rather than in $\sbb$
(~essentially determined by $R_b$). Indeed, if the SM value for the latter
is used, instead of the measured one, to solve for $\abb$ and $\vbb$, the 
discrepancies between the values found and the SM are almost unchanged.
  
  Although the c quark couplings agree well with the SM, and are
  also consistent with the quark-lepton
  universality hypothesis, both $\abb$ and $\vbb$ differ from the SM values
  by more than three standard deviations. The errors on these quantities are,
  however, highly correlated.The statistical significance of these deviations
  is discussed in detail below.
\begin{figure}[htbp]
\begin{center}\hspace*{-0.5cm}\mbox{
\epsfysize10.0cm\epsffile{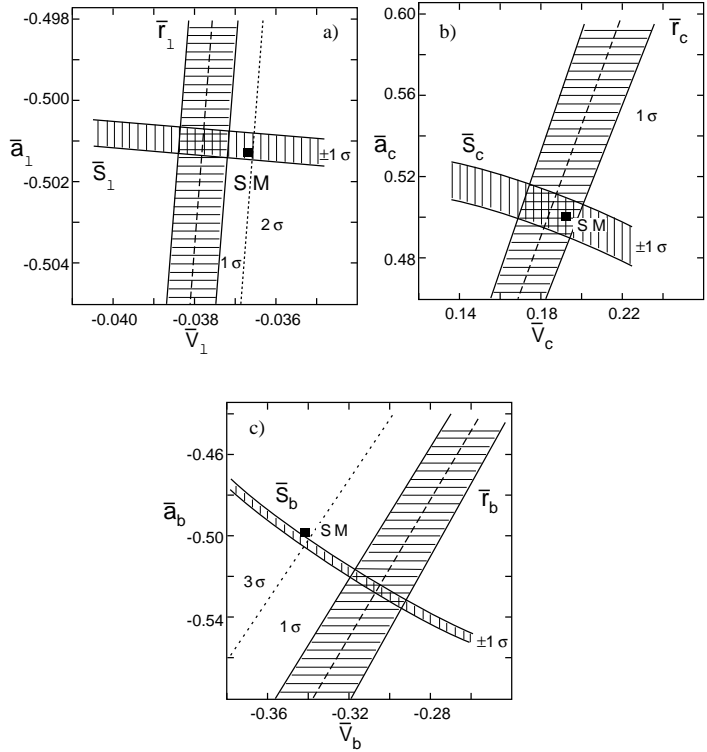}}
\caption{ Constraints on the effective couplings 
$\abf$, $\vbf$ provided by the measurements of $\rbf$ and $\sbf$ .
a) leptons, b) c quarks, c) b quarks. The cross-hatched areas show
$\pm 1 \sigma$ limits. The dotted lines in a),[c)] show $2 \sigma$, 
[$3 \sigma$] limits for $\rbl$,[$\rbb$].
 SM is the Standard Model 
prediction for $m_t =$ 172 GeV, $m_H =$ 149 GeV.}
\label{fig-fig1}
\end{center}
 \end{figure} 
\begin{figure}[htbp]
\begin{center}\hspace*{-0.5cm}\mbox{
\epsfysize10.0cm\epsffile{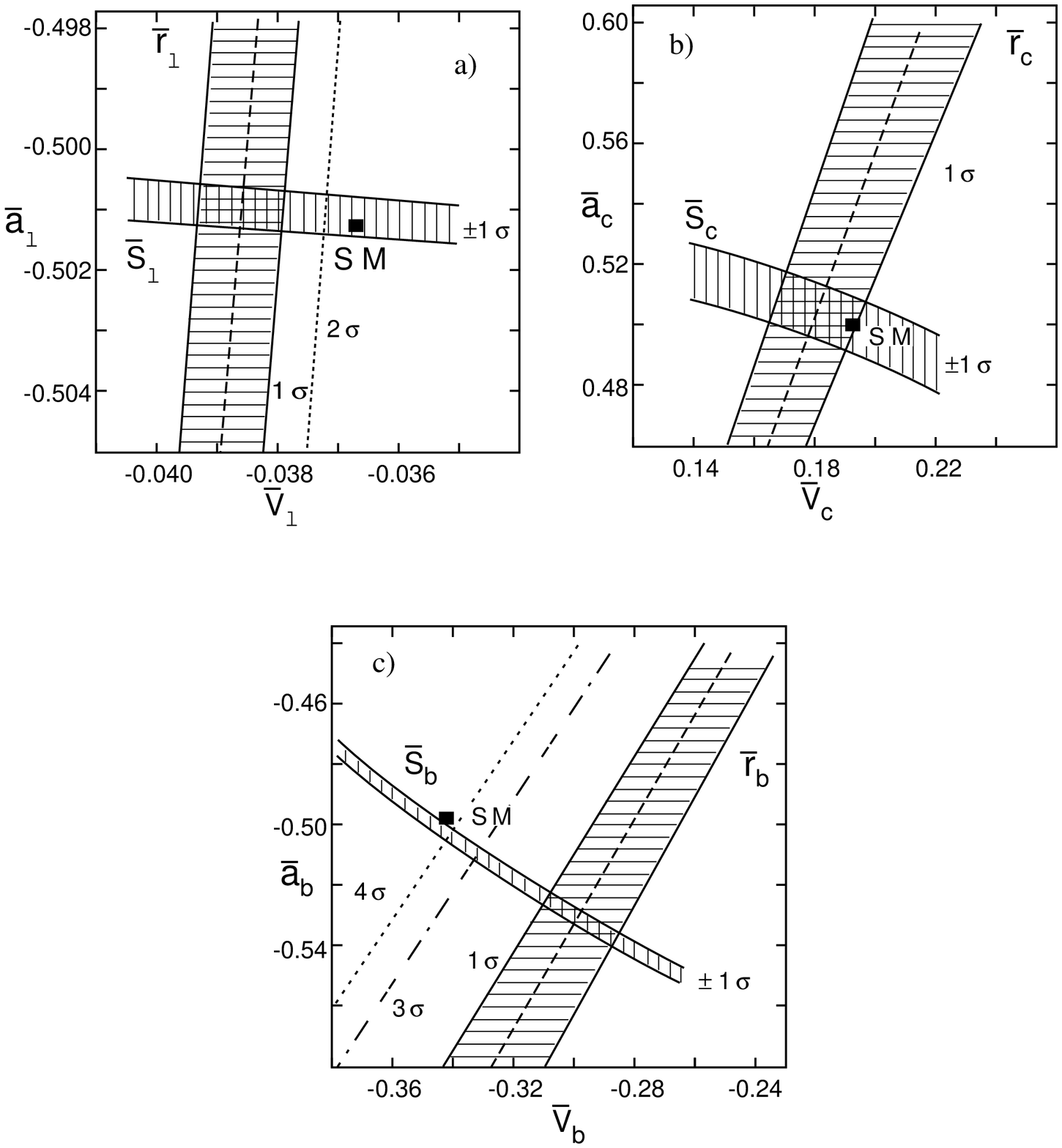}}
\caption{As Fig 1, except that \tpol~ measurements are excluded 
from the LEP average value of $A_l$. The dotted line in a) shows
the  $2 \sigma$ limit for $\rbl$. The dash-dotted [dotted] line in c) shows
 the  $3 \sigma$, [$4 \sigma$] limit for $\rbb$.} 
\label{fig-fig2}
\end{center}
 \end{figure} 

  \par It should be remarked that, although a particular value (0.12) of
   $\alpha_s(M_Z)$
  has been assumed in order to extract the effective couplings of the heavy
  quarks, the sensitivity to the chosen value is very weak. Varying
  $\alpha_s(M_Z)$ over the range $0.1 <  \alpha_s(M_Z) < 0.14$ leads 
  variations of only $\simeq 3 \times 10^{-4}$ in $\abb$ and $\vbb$ to be
   compared with experimental errors $\simeq 1-4 \times 10^{-2}$ (see Table 7).
\par A further constraint on the quark couplings is provided by the
 measurement of the mean quark forward/backward charge asymmetry:
 \begin{equation}
 \langle A_{FB}^q \rangle = \frac{8 A_l \sum_{q} \vbq \abq}{\sum_{q}[(1-6 \mu_q)
 (\abq)^2+(\vbq)^2]}
 \end{equation}
  All experimental analyses performed to date have assumed the correctness 
  of the SM and have used measurements of $\langle A_{FB}^q \rangle$ to
  determine a value of $\sin^2 \theta^{lept}_{eff}$~\cite{x1}. Inserting the
  average value of the latter reported in Ref.[1] into the SM formula
  for $\langle A_{FB}^q \rangle$ and propagating the error leads to the
  `measured' value:
  \[ \langle A_{FB}^q \rangle = 0.1592(86)  \]
  As shown in Table 8 this value is consistent with the SM prediction, 
  with the `model independent' prediction given by inserting the lepton and
 b quark couplings from Tables 5 and 7 into Eqn(2.13)
 and assuming non-b quark lepton universality for the u,d,s,c quarks,
   as well as the
   prediction when, in the latter case, the measured b quark couplings are
   replaced by the SM ones. With the present experimental errors, 
   $\langle A_{FB}^q \rangle$ is therefore insensitive to possible deviations
   of the b quark couplings from the SM, of the magnitude observed in the
   $A_b$ measurements. 
\par As mentioned earlier, in order to avoid having to introduce an accurate 
value of $\alps$ as a correlated parameter in the extraction of the
heavy quark effective 
couplings, the hypothesis of non-b quark lepton universality was made in 
deriving the value of $\sbb$ from the measured quantity $R_b$. The consistency
of this assumption may be checked by extracting $\alps$ from the LEP average value
of $R_l \equiv \Gamma_{had}/\Gamma_l$~\cite{x1}:
\[ R_l = 20.778(29)    \]
using the relation:
\begin{equation}
R_l= 3 \frac{<C^{QED}_q> <C^{QCD}_q>}{C^{QED}_l}\frac{\sum_{q} \sbq}{\sbl}.
\end{equation}
The QED and QCD correction factors $<C^{QED}_q>$ and $<C^{QCD}_q>$
 are averaged over all quark flavours.
The QED correction factors are:
\[ <C^{QED}_q> = 1.00040,~~~~~C^{QED}_l = 1.0019  \]
Inserting the measured values of $\sbb$ and $\sbl$,and using non-b quark lepton
universality to evaluate $\sbq$ ( q $\ne$ b )
gives, for the QCD correction factor:
 \[ <C^{QCD}_q>  = 1.0394(21)    \]
 Using the third order perturbative QCD formula~\cite{x8}:
 \begin{equation} 
 <C^{QCD}_q>  = 1+1.06 \frac{\alps}{\pi}+0.9 (\frac{\alps}{\pi})^2
 -15(\frac{\alps}{\pi})^3
 \end{equation} 
 gives:
 \[   \alps = 0.116_{-0.007}^{+0.005}  \]
 which may be compared to the global fit value of Ref.[1]:
 \[   \alps = 0.120(3)  \]
 The good agreement of the  model independent analysis  result with the
 global world average value: $\alps = 0.118(5)$ found in two recent reviews
 ~\cite{x9,x10} of all published measurements of $\alpha_s$, shows that
 an analysis assuming this value of $\alps$, but without the assumption of
 non-b quark lepton universality, would lead to essentially the same values
 of the b quark couplings as those reported in Table 7.
 In the fit used in Ref.[7]
 to determine the heavy quark effective couplings the constraint
 $\alps = 0.123(6)$ was imposed. As mentioned above, the fitted heavy quark couplings
 are very consistent with those found in the present analysis.    
 \par In order to correctly calculate the statistical significance of the
 deviations from the SM predictions of the effective couplings shown in 
 Tables 5 and 7 it is necessary to take into account the correlations between
 the errors of the different quantities. To avoid the very large correlations
 between the errors on $\abf$ and $\vbf$ (for the case of b quarks the 
 correlation coefficient is -0.96) it is convenient to use, in calculating the
 $\chi^2$, the equivalent quantities $\rbf$, $\sbf$ for which the
 errors are uncorrelated
 for a given fermion flavour $f$. Important correlations still exist, however, 
 between the errors on ($\rbl$, $\rbc$) and ($\rbl$, $\rbb$) in the case
 that $\rbc$ and $\rbb$ are extracted from forward/backward asymmetries
 using Eqns.(2.3),(2.4) and (2.6). The correlation coefficient is:
 \begin{equation}
 {\cal C}_{lQ} = -\frac{(1-\rbl^2)(1+\rbQ^2)}{(1+\rbl^2)(1-\rbQ^2)}
 \frac{\sigma_{\rbl}}{\rbl}\frac{\rbQ}{\sigma_{\rbQ}},~~~(Q=c,b) 
 \end{equation}
 Substituting the parameters from Table 3 gives:
 \[ {\cal C}_{lc} = -0.29,~~~~ {\cal C}_{lb} = -0.52 \]
 The results on the CLs for the agreement with the SM
 of different sets of effective
weak coupling constants, parameterised in terms of $\rbf$ and $\sbf$,
 are collected in Table 9.
These CLs assume perfect statistical consistency of the different
 measurements contributing to the averages. 
The entries in the first column
 of Table 9, giving the level of agreement of ($\rbl$, $\sbl$) with the SM 
 prediction are simply calculated from the entries of Tables 3 and 4
 using a diagonal error matrix, since the errors on $\rbl$ and $\sbl$
 are uncorrelated. 
  Calculating separately the contributions to $\chi^2$ from $\rbl$ and $\rbb$,
 where the latter is derived from the LEP $A_{FB}^{0,b}$ measurement, and
 $\rbb$ derived via Eqn.(2.6) directly from the SLD $A_b$ measurement, 
 gives the entries reported in the second column of Table 9. The CL for
 agreement with the SM prediction of 1.4\% drops to only 0.06\% if the
 \tpol~ measurements of $A_l$ are excluded. The third column of Table 9 results from adding to the
 $\chi^2$ in the second column the (uncorrelated)
 \footnote{Actually there is a weak correlation between $\sbb$ and $\rbl$
 following from Eqn.(2.8), where $\rbl$ is used to calculate $S_Q$. However
 the correlation coefficient is only $\simeq  0.08$ and is neglected here.}
 contributions of $\sbl$ and $\sbb$. In the fourth column of Table 9 the 
 $\chi^2$ and CL of the variables $\rbl$, $\rbb$ and $\rbc$ taking into
 account the $\rbl$-$\rbb$ and $\rbl$-$\rbc$ correlations are given. In the
 last column of Table 9 the (uncorrelated) variables $\sbl$, $\sbb$ and $\sbc$
 are added to those of the fourth column. Note that the number of degrees
 of freedom corresponding to the $\chi^2$ values reported in the 
 second, third, fourth 
 and fifth columns of Table 9 are 3, 5, 5 and 8 respectively, since the 
 $\rbc$ and $\rbb$ measurements derived from the SLD $A_c$, $A_b$ determinations
 give separate, uncorrelated, contributions to the $\chi^2$. 
   As expected, the agreement with the SM improves as the number of degrees
  of freedom of the $\chi^2$ increases (the more parameters are considered,
  the more likely is a deviation associated with any one of the parameters
  to be consistent with a statistical fluctuation). However, there is still
  a factor $\simeq$ 10 difference between the CLs including (or excluding)
  the \tpol~data.
  Taking into account the CL (8.4$\%$) for self-
  consistency of the different $A_l$ measurements, the 
  probability\footnote{Here the term  `probability' is used in the
  the usual sense of the fraction of all cases expected to have
  a CL less than the observed value. For independent $\chi^2$ tests
   the probablities are assumed to be uncorrelated} that all six effective
  couplings are consistent with lepton universality and the SM is 0.9$\%$.
  The similar probability for the leptonic and b quark couplings alone is 0.18$\%$.
 If the \tpol~ measurements are excluded the latter 
  probability drops to 0.018$\%$.
  \par It is important emphasise that a correct calculation of $\chi^2$
  and the associated confidence levels requires that all relevant 
  correlations between errors are taken into account. If a $\chi^2$ is 
  calculated from `raw' experimental measurements, such as those presented
  in Table 1, erroneous conclusions as to the consistency of the data with
  the SM will be drawn. Assuming non-b quark lepton universality,
   the 14 measured electroweak observables presented
  in Table 1 depend on only four unknown parameters, the effective
  couplings of the leptons and of the b quarks. The `raw'
  $\chi^2$ calculated from the `pulls'~\cite{x1}
  in the last column of Table 1 is 21.3 for 14 $dof$ (CL = 0.093 ) or,
  excluding the \tpol~data 19.5 for 12 $dof$\footnote{ In Ref[1], the SM
  prediction is obtained by fitting $m_t$, $m_H$ and several other electroweak
  parameters to the observables of Table 1 as well as others which are not
  directly sensitive to the effective couplings (see Table 20 of Ref.[1]).
  Here, for comparison purposes, the values $m_t =$ 172 GeV, $m_H =$ 149 GeV
  are assumed so that the SM prediction has no free parameters} (CL = 0.077 ).
  The strong sensitivity of the CL to inclusion or exclusion of the
  \tpol~data is completely lost using the `raw' $\chi^2$.
  In fact the $\simeq 2 \sigma$ effects seen in the `pulls' of the observables
  $A_{FB}^{0,\tau}$, $A_e$ (SLD), $A_{FB}^{0,b}$ and $A_b$ (SLD) add, because
  of correlations, constructively in the parameter $\rbb$ to give the 
  observed deviation from the SM of $> 3 \sigma$. First extracting the 
  essential theoretical parameters (the effective couplings), and then comparing
  with the SM predictions using a $\chi^2$ test, is more sensitive to deviations
  of these parameters from the SM predictions than the `raw' $\chi^2$. For the
  latter an inevitable statistical dilution occurs because of the large number
  of experimental observables (14) used as compared to only four effective
  coupling constants that determine the theoretical prediction. This dilution
  effect becomes even more marked when additional observables, not directly
  sensitive to the effective couplings, are added to the $\chi^2$ as in the
  global fit of Ref.[1]. In fact, the five additional observables used in the fit
  contribute only 0.55 to the total $\chi^2$ of 19.1, indicating an overestimation
  of the errors on these quantities\footnote{A $\chi^2$ of 0.55
 for 5 degrees of freedom
 correponds to a CL of 0.990} that reinforces the statistical dilution.
       
\SECTION{\bf{Discussion of the Experimental Measurements of
Electroweak Observables}}
 In order to derive the average values of the electroweak observables presented
 in Tables 1 and 2, many different experimental measurements were combined
 ~\cite{x1}. In the light of the apparent deviations seen from the SM
  predictions in the model independent analysis described in the previous 
  Section, the first question that should be asked is whether the 
  experimental measurements are reliable and consistent. An important
  general question is whether the uncertainty on the measured observable is
  dominated by statistical or systematic errors. Only in
  the former case, the error can be interpreted, with confidence, in the
  statistical sense ($1 \sigma \equiv 68\%$ CL) and the
  probabilistic meaning of the
   CL of a $\chi^2$ test can be expected to be reliable.
  This is no longer
  the case if the systematic error is dominant. As there is no definite,
   agreed, procedure for assigning systematic errors, the meaning of the 
   error can depend on psychological (or even sociological) factors.
   If the physicist is over-conservative in assigning the error, real 
   deviations from a theoretical expectation can be missed, or in the 
   contrary case spurious detector related effects wrongly interpreted
   as `new physics'. Some check on the degree of conservatism, or
    otherwise, of physicists is however provided if there are many repeated
    measurements of the same quantity, 
    as provided, in the present case, by the different
    experiments at LEP. When the errors are dominated by 
    systematic effects, or contain a large systematic contribution, then 
    comparing the weighted average error to that calculated by applying
    the Central Limit Theorem to the different measurements of the
    same quantity gives an indication whether the systematic errors
    are over- or under-estimated. Such a test only applies to errors which
    are uncorrelated for the different measurements of the same
    quantity. This test may be best applied when each experiment measures
    the same observable using a similar method. If consistency is found
    in this case, but inconsistencies are found when the same physical
    quantity is measured using different observables ( for example,
    the quantity $A_l$ may be measured using either forward/backward 
    asymmetries for different dilepton final states, or \tpol~) it is probable
    that there is an unknown source of correlated systematic error.
    As discussed further below,
    such an error can arise, for example, due to an inadequate treatment of
    QED radiative corrections.  
    \par The apparent deviations from the SM predictions of the b quark
    couplings seen in Table 7 are due those observed in only two of the
    basic electroweak parameters: $A_b$ and $A_l$. Most of the current 
    information on $A_b$ is derived from the Z-peak $A_{FB}^{0,b}$ 
    measurements of the four LEP experiments. The error on this $A_b$
    measurement is roughly half that of the SLD $A_b$ measurement.
     The individual measurements of the experiments~\cite{x1} 
     contributing to the LEP average $A_{FB}^{0,b}$ measurement quoted
     in Table 1 are shown in Fig. 3, together with the weighted average
     value, its error, and the SM prediction. There is no hint of
     any badly understood systematic effect in the distribution of these
     measurements. Except for the ALEPH jet-charge measurement~\cite{x11},
     all errors, even those using all LEP1 data, are statistically
     dominated. The weighted average value agrees well with the individual
     measurements ($\chi^2 = 5.9$ for 8 $dof$, CL$ = 66\%$). 
     The estimate of the error on the mean value, given by
     the sample variance of the seven most accurate determinations,
     using the Central Limit Theorem is 0.00232, in excellent agreement
     with the weighted average error of the same data points, which is
     0.00237. The situation is very different for the parameter $A_l$.
     Shown in Fig. 4 are the values of $A_l$ derived from the LEP
     measurememts of $A_{FB}^{0,e}$, $A_{FB}^{0,\mu}$, $A_{FB}^{0,\tau}$,
     $A_e$ and $A_{\tau}$ from \tpol~ and $A_e$ as measured by SLD 
     using the left/right beam polarisation asymmetry. Although the overall 
     consistency of the individual measurements with the weighted 
     average value seems acceptable ($\chi^2 = 9.7$ for 5 $dof$, CL$ = 8.4\%$)
     the internal consistency of the various measurements is much worse.
     In particular there are three, essentially independent,
     \footnote{There is a weak correlation between two of the three effects
     ,in that the $A_{FB}^{0,e}$, $A_{FB}^{0,\mu}$ measurements, used in
     the first consistency check, contribute also to the `non-$\tau$' weighted
     average value of $A_l$ used in the second. This can be avoided by
     comparing the \tpol~value of $A_l$ with the $A_{LR}$ measurement.
     In this case an even larger discrepancy of 2.4$\sigma$ is found.}
     $\simeq 2-3\sigma$ deviations concerning $\tau$-related
     measurements:
     \begin{itemize}
     \item $A_{FB}^{0,\tau}$ is 1.8$\sigma$ higher than the average of 
      $A_{FB}^{0,e}$ and $A_{FB}^{0,\mu}$. Also, in all 4 LEP experiments,
      (see Table 3 of Ref.[1]) $A_{FB}^{0,\tau}$ is higher than
        $A_{FB}^{0,e}$ or $A_{FB}^{0,\mu}$. Assuming no systematic bias,
        the probability for this is 1 in 81.
     \item The average value of $A_l$ extracted from the \tpol~data:
     0.1393(50) lies 2.2$\sigma$ below that, 0.1522(30), given by the
     weighted average of the non-$\tau$ measurements.
     \item Assuming lepton universality, the LEP average value of
      $A_{FB}^{0,\tau}$ gives, using Eqn.(2.3), $A_l = 0.1649(71)$. This is
      2.9$\sigma$ higher than the mean $A_l$ calculated from the 
      \tpol~ measurements of $A_e$ and $A_{\tau}$.
      \end{itemize}
      These deviations cannot be explained by a breakdown of lepton 
      universality for the $\tau$. Using the \tpol~$A_e$ measurement to 
      extract,using Eqn.(2.3), $A_{\tau}$ from $A_{FB}^{0,\tau}$ gives
      $A_{\tau} = 0.1968(204)$. This may be compared with the 
      \tpol~ measurement: $A_{\tau} =$ 0.1401(67). There is a 2.6$\sigma$
      discrepancy. In the case of a breakdown of lepton universality, the two
      determinations of $A_{\tau}$ must give a consistent result that is 
      significantly different from the measured $A_{e}$ value. In fact, the
      values of $A_{e}$ and $A_{\tau}$ found using \tpol~ are 
      consistent within 0.19$\sigma$.
\begin{figure}[htbp]
\begin{center}\hspace*{-0.5cm}\mbox{
\epsfysize10.0cm\epsffile{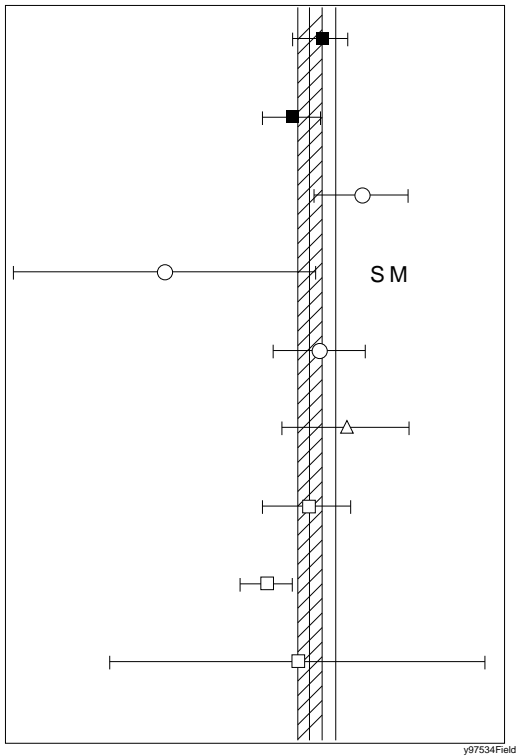}}
\caption{ LEP on-peak $A_{FB}^{0,b}$ measurements. ALEPH: solid squares,
DELPHI: open circles, L3: open triangle, OPAL: open squares. The $\pm 1
\sigma$ region around the weighted average value is indicated by the 
hatched band. The vertical line is the Standard Model prediction for
 $m_t =$ 172 GeV, $m_H =$ 149 GeV.}
\label{fig-fig3}
\end{center}
 \end{figure} 
\begin{figure}[htbp]
\begin{center}\hspace*{-0.5cm}\mbox{
\epsfysize10.0cm\epsffile{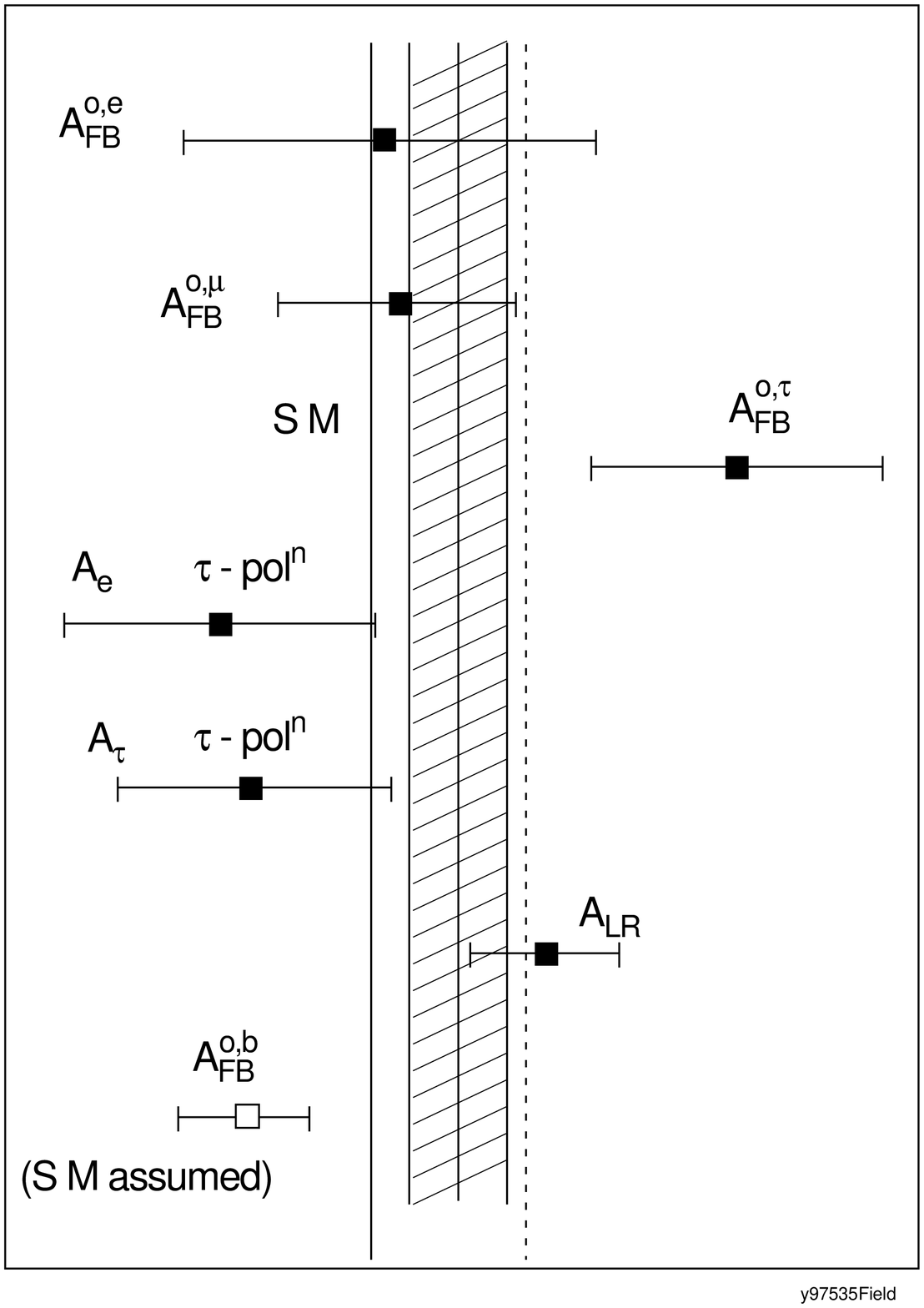}}
\caption{ LEP and SLD $A_l$ measurements. The hatched band shows
the $\pm 1\sigma$ region around the weighted average value. The weighted 
average value, excluding the \tpol~ measurements, is given by the 
dashed vertical line. The solid vertical line is the 
Standard Model prediction for
 $m_t =$ 172 GeV, $m_H =$ 149 GeV. The open square shows the value of
 $A_l$ derived from the LEP average value of $A_{FB}^{0,b}$ assuming the SM;
 this datum is not included in the weighted averages shown.}
\label{fig-fig4}
\end{center}
 \end{figure} 
    \par Assuming gaussian errors, the probability that {\it all} these 
    $\tau$-related 
    apparent deviations from lepton universality are statistical
    fluctuations, is $7.5 \times 10^{-6}$. This situation may be compared with
     that for the non $\tau$-related measurements of $A_l$ derived from
     $A_{FB}^{0,e}$, $A_{FB}^{0,\mu}$ and $A_{LR}$ which are, respectively
      $0.147(11)$,~~$0.148(61)$ and $0.154(4)$. Labelling these measurements
      1,2,3 respectively, the deviations between the pairs 1-2, 1-3 and 2-3 are
      0.08$\sigma$, 0.6$\sigma$ and 0.83$\sigma$. The measurements are perfectly
      consistent. 
    \par It may be remarked that the $A_{FB}^{0,\tau}$ measurement,
     which occurs in the first of the three above-mentioned 
     apparent deviations, has been
     included in the evaluation of the average values of $A_l$ used in the
     model-independent analysis of the previous Section. However, excluding
     this datum from the $A_l$ average in addition to the \tpol~data gives
     $A_l = 0.1516(29)$. Thus, (see Table 2) the deviations from the SM 
     predictions will lie between the `\tpol~ in' and `\tpol~ out' cases
     discussed above, if all $\tau$-related measurements are excluded.
     \par Also shown in Fig. 4 is the value of $A_l$ derived from 
$A_{FB}^{0,b}$, assuming the correctness of the SM. The value so obtained,
0.1396(33), differs from the weighted average of the purely leptonic
 measurements by 2.6$\sigma$, or by 3.2$\sigma$ if the \tpol~data are
 excluded. It is clear, from the analysis of the previous Section, that 
 these discrepancies are mainly due to the deviations of the b quark effective
 couplings from the SM predictions. The quantity $\sin^2\theta_{eff}^{lept}$
 used in Ref.[1] is directly related to $A_l$ via Eqns.(2.4),(2.12). The poor
 consistency of the different $\sin^2\theta_{eff}^{lept}$ determinations
 in Table 19 of Ref.[1] is largely due to
 the inclusion of values derived from $A_{FB}^{0,b}$ and 
 $\langle A_{FB}^q \rangle$ assuming the correctness of the SM. The common 
 origin, in the b quark couplings, of the poor agreement of the different
  $\sin^2\theta_{eff}^{lept}$ determinations and the 3$\sigma$ deviation of
  the measured LEP-SLD average value of $A_b$ from the 
  SM prediction, was not pointed out in
  Ref.[1].
  \par Each type of observable contributing to the average values of $A_l$:
   $A_{FB}^{0,l}$, ($l=e,\mu,\tau$) $A_e$ and $A_{\tau}$ from \tpol~ and 
   $A_e$ from $A_{LR}$ are now discussed it turn, taking into account the
   internal consistency of measurements of the same quantity performed by
   different experiments, the relative importance of statistical and
   the estimated systematic errors, and possible systematic effects 
   that may not have, so far, been taken into account.
   
   \par \underline{{\it The LEP Leptonic Forward/Backward
    Charge Asymmetry Measurements}}
   \newline
   The values of $A_{FB}^{0,l}$, ($l=e,\mu,\tau$) for the different LEP 
   experiments are presented in Table 3 of Ref.[1]. A detailed breakdown 
   of the systematic errors of the different
   data sets is found in Table 2 of the 
   same paper. Although the individual experimental arrors are usually
   statistics dominated, the statistical and systematic contributions
   to the error on the weighted average (see Table 20
   \footnote{Note there is a mis-print in this Table. The estimated 
   systematic error on $A_{FB}^{0,l}$ should presumably be 0.0007
   not 0.007.} of Ref.[1]) are almost equal. The most remarkable systematic
   feature of the $A_{FB}^{0,l}$ measurements, as already mentioned above, is
   the relatively high value of $A_{FB}^{0,\tau}$ found by all four
   experiments. Since each of the LEP experiments have comparable statistics in
   each channel, each can be considered to provide an independent estimate,
   with a similar weight, of $A_{FB}^{0,l}$. Disregarding the estimated errors
   on each measurement, an unbiased statistical test of the consistency of
   the measured values of $A_{FB}^{0,l}$ can be made using the 
   Student's t distribution. This is done by calculating
   the probability that, say, the $A_{FB}^{0,\tau}$ and  $A_{FB}^{0,\mu}$
   measurements of the different experiments are consistent with a common mean
   value~\cite{x12}. The results of this comparison for the 3 possible pairings
   ($\tau$, $\mu$), ($\tau$, $e$) and ($\mu$, $e$) are presented in Table 10. 
    Also shown in this Table are the CLs for consistency of the measurements,
    based on the total errors quoted in Table 1. The systematically larger
    CLs found using these errors is perhaps indicative that the
     point-to-point systematic errors tend to be over-estimated. 
   The good agreement between the $\mu$-$e$ measurements and the poor
   agreement between $\tau$-$\mu$ and  $\tau$-$e$ indicates the possible 
   presence of a correlated systematic effect, not included in the present
   systematic error estimate, for the $\tau$-pair channel.
   \par  An obvious 
   candidate for for such an effect is the QED radiative correction.
   On the Z-peak this is large; the combined O($\alpha$) and  O($\alpha^2$)
   corrections amount to $\simeq -110\%$ of the corrected pole
   asymmetry for muon-pairs~\cite{x13}. The systematic error on the 
   weighted average value of $A_{FB}^{0,l}$ is then $\simeq 4.4\%$ of the
    radiative correction, and the observed deviation of the $\tau$
    from the combined $\mu$-$e$ results is $\simeq 27\%$ of it.
    It seems, however, unlikely that the error on the
    theoretical estimate of the QED radiative correction, essentially due to
    unobserved initial state radiation, and the associated virtual
    corrections, could be large enough to explain the high value of the 
    $\tau$-pair asymmetry. One effect that can produce large changes in
    the forward/backward asymmetry is intial/final state interference
    in the case that hard cuts are applied to the radiated photons~\cite{x14,x15}.
    Indeed, for some decay channels, such as $\pi \nu$, $\pi \pi \pi \nu$, tight
    cuts are applied in some of the LEP analyses in order to cleanly separate
    them from channels containing extra $\pi^0$s. Such effects could be investigated
    by comparing the forward/backward asymmetries for different $\tau$ decay modes
    and different cuts on additional photons.
 \par \underline{{\it The LEP $\tau$-Polarisation Measurements}}
\newline    
Results of the measurements of the average \tpol~ of the four 
LEP experiments~\cite{x16,x17,x18,x19}, for each $\tau$ decay channel analysed, 
are presented in Table 11. The first error shown on each measurement is
statistical, the second one systematic. The weighted average values  $A_{\tau}$
and errors are given separately for each experiment and for each decay channel.
The overall weighted average and its error are also given. The results shown in
Table 11 do not correspond exactly to those used in the LEP averages reported
in Ref.[1], although there is a large overlap. The data sets chosen are those
for which both statistical and systematic errors
 have been given for each decay channel. 
\par It can be seen that, unlike for the forward/backward charge asymmetry
measurements, the quoted statistical and systematic errors of individual
experiments are comparable for almost every decay channel. The systematic error
estimates can be tested by comparing the weighted errors, for each experiment
or each decay channel, with the error estimate on the mean value
given by the Central Limit Theorem: $\sqrt{\sum(x-\overline{x})^2/n(n-1)}$.
The latter error estimates are given in Table 11 in square brackets next to
the weighted averages.  It can be seen that the level of agreement is better
between experiments than between decay channels. For the $e \nu \nu$,
 $\mu \nu \nu$, $\pi \nu$, and $a_1 \nu$  channels
  there is an indication of an over-estimation of
  the systematic errors, while for $\rho \nu$ the error calculated
 from the sample variance is much larger than the weighted average error.
 This is due to the wide spread of the measurements in this channel; the DELPHI
 and L3 values are much larger than those of ALEPH and OPAL
 \footnote{The weighted average of DELPHI and L3 is $A_{\tau} = 0.171(19)$, that
 of ALEPH and OPAL $A_{\tau} = 0.113(15)$. The difference is 2.5 standard
 deviations}.
 \par The common mean test using the Student's t distribution, applied above to the 
 forward/backward charge asymmetry measurements has also been applied to the 
 measurents of $A_{\tau}$ presented in Table 11. The results of the comparisons of different
 decay channels are presented in Table 12 and of different experiments in Table 13.
 In each case the CLs for consistency based on the total experimental errors
 are also given. The main systematic features of the $A_{\tau}$  measurements are
 the relatively
 high values found by all experiments for the $e \nu \nu$ channel and the high
 value of the L3 weighted average, as compared to those of the other three 
 experiments. The latter effect is also seen in the
 \tpol~measurements of $A_e$~\cite{x1}.
  The systematically larger value for the  $e \nu \nu$ channel is reflected in
 the poor confidence levels (0.35$\%$-2.5$\%$) of the
 Student's t tests for the channels $e-\mu$, $e-\pi$  and $e-a_1$ in Table 12.
 Relatively worse CLs are also seen for these 
 channels in the test based on the total errors, but as expected from the larger 
 experimentally assigned errors as compared to those calculated from the sample
 variance, higher absolute CLs (11$\%$-31$\%$) are found. The agreement between
 the two types of test is much better in Table 13, from which one might be 
 tempted to conclude that all 4 experiments give consistent results\footnote{
The Student's t test is less well adapted to the comparison
of the different LEP experiments since the different decay channels have
very different statistical sensitivites, whereas each test datum is treated,
in the test, on
an equal footing. On the other hand, since each experiment has comparable 
statistics and sensitivity for a given decay channel, the  Student's t 
common mean test is well adapted to the channel-by-channel comparisons}.

\par As in the case of the $\tau$ forward/backward charge asymmetry, the radiative
correction associated with final state radiation appears as an obvious 
candidate to explain both the systematic differences observed between 
different decay channels and those, mentioned above, between the \tpol~ and 
other measurements of $A_l$. The final state radiative corrections in the \tpol~
measurements are not only large for most decay channels, but depend strongly on
 the detection efficiency
of the radiated photons and hence on the acceptance and resolution of the 
LEP detectors as well as the experimental cuts. The \tpol~ is measured by fitting
the energy spectra of $\tau$ decay products. The latter are directly effected
by the rate and energy spectra of, and cuts applied to, the radiated photons.
Any systematic errors in the treatment of final state radiation are thus 
directly correlated to systematic errors in the  \tpol~ measurement. Consider,
for example, the measurement of $\langle P_{\tau} \rangle$ using the $\pi \nu$
decay channel on the Z peak. Measuring only the pion energy and neglecting
that of the radiated photons has been estimated~\cite{x20} to shift 
$\langle P_{\tau} \rangle$ by $\simeq$ 20$\%$ of its value. This may be compared
to the systematic error assigned to the LEP average value of  $\langle P_{\tau} \rangle$
of 3.3$\%$ of its value~\cite{x1}. Since radiative corrections are not included in
the systematic error estimate, a tacit assumption is thus made that they are 
known to much better than $\simeq 10\%$ of their value.
\par All of the LEP experiments have used the same Monte Carlo program, 
KORALZ~\cite{x21}, to correct for radiative effects in fitting the energy spectra of
the $\tau$ decay products. No experimental checks (measurements of the rate and
distributions of final state photons) have been published, and any systematic
errors assigned by the experiments for radiative corrections have been very small
in comparison to detector-related sources of systematic error.
As for the case of the $\tau$ forward/backward asymmetry it seems unlikely, however,
that the approximations made in simulating the
 radiative corrections can account for the all the different values of
$A_{\tau}$ found using different decay channels. For the decay mode of
largest statistical power, $\tau \rightarrow \rho \nu$, where the 
sensitivity to the \tpol~ is given by angular information
from the $\rho$ decay as well as the energy distributions of
decay products, the sensitivity to radiative corrections is
small, The entire radiative correction generated by KORALZ has been
shown to generate a shift of only -0.011 in $A_{\tau}$ in this 
case\footnote{W.Lohmann, private communication}.   
 For the case of the $\pi \nu$ decay channel,
a large `structure dependent' effect in the
final state photon spectrum due to the decay $a_1 \rightarrow \pi \gamma$ is to be 
expected. This contribution and its interference with the bremsstrahlung amplitudes,
have been estimated~\cite{x22} to change the yield of high energy photons with 
$x = 2E_{\gamma}/m_{\tau} > 0.6$ by a factor $\ge 2$. However, the radiative correction to $A_{\tau}$
is dominated by soft bremsstrahlung photons, so no large corrections are to be expected from this
 effect. For the $e \nu \nu$, $\mu \nu \nu$ decays KORALZ uses the exact 
O($\alpha$) matrix element. Particularly for the electron case where the 
radiative corrections are large it is perhaps, however, of interest to investigate the effect
of O($\alpha^2$) and higher order corrections.
\par Independently of any specific conjectures on possible sources
of systematic effects, the systematic error
of 0.0045 (3.3$\%$) assigned to the LEP average value of $A_{\tau}$ in Ref.[1] would
 appear to be unduly optimistic, in view of the apparent inconsistencies between
 measurements
 from both different decay channels and different experiments, discussed above. The value
 of $A_{\tau}$ derived from the $e \nu \nu$ channel deviates from the overall weighted 
 average by 8.6 times the above systematic error estimate, and the L3 experiment by
 4.4 times. The Student's t common mean test shows that the
 uncorrelated systematic errors on the
 individual experiments are probably, on average, overestimated. Thus the real 
 inconsistencies between experiments and decay channels are probably larger than those
 estimated from the assigned experimental errors. This argues even more strongly 
 that the assigned systematic error on the average value of $A_{\tau}$ is too small.
 \par To illustrate the effect of an under-estimation of the systematic error in the
 $\tau$-polarisation measurements,
 consider the effect of doubling the systematic error on the LEP
 average value of $A_{\tau}$, and assigning the same systematic error to $A_e$ 
 measured from the $\tau$-polarisation asymmetry\footnote{
 In Ref.[1] a systematic error of only 0.002 is assigned to $A_e$. Why this should 
 be less than half of the already small error assigned to $A_{\tau}$ is not clear to
 this writer, taking into account the close similarity of the two measurements.
 Table 20 of Ref.[1] contains, however, the disclaimer that: `the determination
 of the systematic part of each error is approximate'.}.
 
 This results in a weighted average value of $A_l$ of 0.1518(26), which lies almost mid-way
 between the `\tpol~-in' and `\tpol~-out' values shown in Table 2. 
 \par The conclusion of this discussion of the \tpol~ measurements is that there is evidence 
 that the uncorrelated systematic errors of the different experiments for the same 
 decay channels have been overestimated. On the other hand correlated systematic
 errors between different decay channels seem to have been underestimated.
 The true world-average value of $A_l$ probably
 lies between the two values quoted in Table 2.
 \newline           
 \par \underline{{\it The SLD $A_{LR}$ Measurement}}
\par The discussion of this measurement can be very brief, as there is only one
 experiment and the present error is strongly statistics-dominated~\cite{x23}.
 The relative systematic error on $A_{LR}^0 = A_e$, which is almost completely 
 determined by that on the SLC electron beam polarisation, is only 0.7$\%$,
 as compared to the relative statistical error of 2.8$\%$. The accuracy
 of the beam polarisation
  measurement has been checked at the 4$\%$ level by independent determinations
 using Compton and M\o ller polarimeters. For comparison, the difference
 between the LEP \tpol~ and the
 SLD $A_{LR}$ measurements of $A_l$ amounts to 8$\%$. The systematic error of the 
 $A_{LR}$ measurement would have to be wrong by an order of magnitude to account 
 for this difference. Another very important advantage of the $A_{LR}$ 
 measurement is the low level of the QED radiative
 correction, of only $\simeq -3\times 10^{-4}$
 ~\cite{x24} as compared with -110 $\%$ for $A_{FB}^{0,l}$ or
 -8$\%$ to +20$\%$, depending on the
 decay channel and the cuts on the final state radiation~\cite{x20},
 for $\langle P_{\tau} \rangle$.
 
 \par To summarise the above detailed discussion of the different measurements
 contributing to the world average value of $A_l$ used in the model independent
 analysis of Section 2; although the two LEP measurements ($A_{FB}^{0,l}$  and \tpol~)
 and the SLD $A_{LR}$ measurement have, currently, similar statistical errors
 the situation is very different with respect to systematic errors. Only for the
 $A_{LR}$ measurement is the systematic error negligible. For $A_{FB}^{0,l}$ and
 $\langle P_{\tau} \rangle$ the estimated statistical and systematic errors are
 almost equal~\cite{x1}. However, inconsistencies at the 2-3 standard deviation 
 level in $\tau$-related measurements of both $A_{FB}^{0,l}$ and 
 $\langle P_{\tau} \rangle$ indicate that the true systematic error is probably 
 considerably larger. 
 
\SECTION{\bf{Physical Interpretation of the Measured Effective 
Weak Coupling Constants}}
 The test of the SM provided by measurements of Z decays
at LEP and SLD is, essentially, that
of the SM predictions for the quantum corrections, arising from 
massive virtual particle loops, to the Born level diagrams for $e^+e^- \rightarrow
 Z \rightarrow f \overline{f}$.
These corrections may be conveniently expressed in terms of two parameters
$\Delta \rho_f$ and $\Delta \kappa_f$ for each fermion flavour~\cite{x25}. The
 parameters
 are given, in terms of the effective couplings, by the relations:
\begin{eqnarray}
\Delta \rho_f & = & -2(1-2|\abf|)  \\
\Delta \kappa_l & = & \frac{(1-\rbl)}{4 s_W^2}-1  \\
\Delta \kappa_c & = & 3\frac{(1-\rbc)}{8 s_W^2}-1  \\
\Delta \kappa_b & = & 3\frac{(1-\rbb)}{4 s_W^2}-1
\end{eqnarray}
Here, following the usual on-shell definition~\cite{x26}:
\begin{equation}
\sin^2 \theta_W = s_W^2 = 1-c_W^2 \equiv 1- \frac{M_W^2}{M_Z^2}
\end{equation}
Since $\Delta \kappa_f$ is determined by $\rbf$, only the weak theoretical assumption of
lepton universality is needed to extract it from the experimental measurements.
\par  The SM predictions of Section 2 used the fixed values:
$m_t = $ 172 GeV, $m_H = $ 149 GeV
found in the global fit of Ref.[1].
The effect on the SM prediction of varying $m_t$ and $m_H$ within the
existing experimental bounds~\cite{x27,x28} is now considered. The dependence
of $\Delta \rho_f$ on $m_t$ and $m_H$ is contained in the terms~\cite{x25}:
\begin{eqnarray}
\Delta \rho_f^{top} & = & \frac{ 3 G_{\mu} m_t^2}{8 \sqrt{2}\pi^2}(1+\xi_f) \\
\Delta \rho_f^{Higgs} & = &    -\frac{\sqrt{2} G_{\mu} M_W^2}{8 \pi^2}
\tan^2 \theta_W
\left[ \frac{11}{3}\left(\ln \left(\frac{m_H}{M_W}\right)-\frac{5}{12}\right)\right] 
\end{eqnarray}
where $\xi_f = 0$ for $f \ne b$ and -4/3 for  $f = b$. The quantum correction
$\Delta \kappa_f$ is calculated using a parameterisation\footnote{The 
relative accuracy of the formula (4.8) is about one per mille for the
interesting range of values of  $m_t$ and $m_H$.} of the ZFITTER~\cite{x29}
 prediction of the effective leptonic weak mixing angle:
\begin{equation}
(\overline{s}_W^l)^2 = 0.233597-8.95 \times 10^{-8}m_t^2-3.86 \times 10^{-4}
\ln m_t+5.43 \times 10^{-4}\ln m_H
\end{equation}
where  $m_t$ and $m_H$ are in GeV units. $\Delta \kappa_f$ is related to
$(\overline{s}_W^l)^2 = (1-\rbl)/4$ by Eqns. (2.1),(2.9),(2.10) and (4.2-4.4). For
the b quark there is an additional non-universal contribution:
\begin{equation}
\Delta \kappa_b^{top}  =  \frac{  G_{\mu} m_t^2}{4 \sqrt{2}\pi^2}.
\end{equation}

\par The values of $\Delta \rho_f$, $\Delta \kappa_f$ for $f = l,~c,~b$,
extracted from the measured
effective couplings using Eqns.(4.1)-(4.5), are presented in Table 14.
Standard Model predictions are shown for the cases $m_t = $ 172 GeV,
 $m_H =$ 149 GeV and for $m_t = $ 180 GeV, $m_H =$ 100 GeV. The latter
choice gives a somewhat better description of the leptonic corrections.
 In Table 15 corresponding results for $\Delta \rho_f$ and $\Delta \kappa_f$
 are shown for the case when the \tpol~ measurements are excluded from 
 the LEP average value of $A_l$.
\par  Good agreement with the SM is seen for leptons and c quarks.  For b quarks however,
the measured values of the quantum corrections are much larger than the SM
predictions. For $\Delta \rho_b$ the measured value exceeds the SM prediction
by a factor of 13-15, and is of opposite sign. The measured value of $\Delta \kappa_b$
has the same sign as the SM prediction, but is 9-11 times larger. Both effects are at
the $>$ 3 standard deviation level, but they are highly correlated. 
The discrepancies seen are
so large that the significance of the deviations shows almost
no sensitivity to $m_t$ and $m_H$.     
\par It is also instructive to present the quantum corrections in terms of the
`epsilon parameters' introduced by Altarelli et al.~\cite{x30,x31,x32}. In terms of 
the variables used in the present paper to describe the effective couplings, these
are defined as~\cite{x30}:
\begin{eqnarray}
\epsilon_1 & \equiv & \Delta \rho_l = -2(1+2\abl)  \\
\epsilon_2 & \equiv & c_0^2 \Delta \rho_l+
\frac{s_0^2 \Delta r_W }{(c_0^2-s_0^2)}-2 s_0^2 \Delta k' \\
\epsilon_3 & \equiv & c_0^2 \Delta \rho_l+(c_0^2-s_0^2) \Delta k'
\end{eqnarray}
here $s_0^2 = 1-c_0^2$ and $\Delta r_W$ are defined by the relations:
\[ \left(1-\frac{M_W^2}{M_Z^2}\right)\frac{M_W^2}{M_Z^2} = \frac{s_0^2 c_0^2}{1-\Delta r_W}
= \frac{\pi \alpha(M_Z)}{\sqrt{2} G_{\mu} M_Z^2(1-\Delta r_W)} \]
and  
\[ \Delta k' = \frac{(1-\rbl)}{4 s_0^2} -1. \]
In Ref.[32] a fourth parameter, $\epsilon_b$, was introduced. It may be
 defined in three distinct ways:
\begin{eqnarray} 
\epsilon_b(\abb) & \equiv & \frac{\abb}{\abl}-1   \\
\epsilon_b(\rbb) & \equiv & \frac{\rbb-{\cal R}_l}{1-\rbb}  \\
\epsilon_b(\sbb) & \equiv & \frac{\sbb-(\abl)^2(1-6 \mu_b+{\cal R}_l^2)}
{2 (\abl)^2 (1- 6 \mu_b+2{\cal R}_l)}
\end{eqnarray}
where
\[ {\cal R}_l = \frac{(2+\rbl)}{3}.  \]
In the SM, retaining only the leading terms $\simeq  m_t^2$, the three
definitions (4.13)-(4.15) are equivalent\footnote{ Modulo small b-mass
dependent corrections}. In previous phenomenological applications
however,~\cite{x32,x33} only the third definition (4.15) based, via Eqn.(2.8)
on the measurement of $R_b$ was used. The measured values of the
six epsilon parameters defined above are presented in Table 16, where 
they are compared with the SM predictions. 
 As noted previously~\cite{x33} the values of
$\epsilon_1$, $\epsilon_2$ and $\epsilon_3$ are in good agreement with
the SM predictions. A small deviation is observed for $\epsilon_b(\sbb)$, a
residual of the much discussed~\cite{x1} `$R_b$ problem'. However,
both $\epsilon_b(\abb)$ and $\epsilon_b(\rbb)$ deviate from the SM
prediction by about four standard deviations\footnote{Again, the errors
on these quantities are highly correlated}.
 One may note the extreme sensitivity of the parameter $\epsilon_b(\rbb)$
 to the anomalous b coupling; the measured value is 39 times
 and 4.7$\sigma$\footnote{The errors on this quantity, determined essentially
by those on $A_b$, are skewed and non-gaussian. The average error is quoted
in Tables 16 and 17. The confidence level of the deviation of
 $\epsilon_b(\rbb)$ from the SM, assuming gaussian errors for $A_b$,
 is in fact almost the same as that of the latter, about one per mille.}
 larger than the
 SM prediction. The SM predictions for
these quantities are insensitive to $m_H$ and are essentially
given by the term $\simeq m_t^2$:
\begin{equation}
   \epsilon_b = -\frac{2}{3} \Delta \rho^{top} = -\frac{G_{\mu} m_t^2}
   {4\sqrt{2} \pi^2} = -0.0062~~~(m_t = 172 {\rm GeV})
\end{equation}
Table 17 shows the epsilon parameters calculated excluding the \tpol~ measurements
from the averages. 
It can be seen that the measured value of $\epsilon_b(\rbb)$ exceeds the SM 
prediction by a factor of 44 and more than six standard deviations in this case.
It may be remarked that, in this case, the leptonic parameter $\epsilon_3$ also shows a deviation
of 3 to 4$\sigma$.  
\par The conclusion to be drawn from Tables 16 and 17 is that the deviations 
observed for the b quark couplings, interpreted as a real physical 
effect, do not enter at all into the framework of the SM nor any of
its `natural' extensions. Supersymmetry, Technicolour, anomalous
$WW\gamma$ or $WWZ$ couplings, and new $U(1)$ gauge bosons are all
expected, via vacuum polarisation effects in the gauge boson 
propagators, to produce deviations from the SM predictions for
 $\epsilon_1$, $\epsilon_2$
 or $\epsilon_3$~\cite{x30,x31,x32}. These parameters are much more
sensitive to any flavour independent modifications of the
couplings, due to anomalous vacuum polarisation effects, because of
the high precision of the purely leptonic measurements.
  Although the  $\epsilon_3$ parameter shows a quite large deviation 
from the SM prediction in the case that the \tpol~ measurements
are excluded, the most important apparently anomalous effect occurs in 
 the quantum corrections to the 
 b quark couplings that disagree, by an order of magnitude, with the
 expectations of the SM.
 \par A clue as to the origin of the anomalous b quark couplings is
  provided
 by considering the right- and left-handed effective couplings, $\gbr$, $\gbl$ related to
 $\abb$ and $\vbb$ by the relations:
\begin{eqnarray}
\gbr = \frac{1}{2}(\vbb-\abb) & = & -\sqrt{\rho_b} e_b (\overline{s}_W^b)^2 \\
\gbl = \frac{1}{2}(\vbb+\abb) & = & \sqrt{\rho_b}[T_b^3- 
e_b (\overline{s}_W^b)^2 ]
\end{eqnarray}
From the measured values of $\abb$ and $\vbb$ presented in Table 7, the following  
values of the left-handed and right-handed effective couplings of the b quarks are
found:
\[ \gbl  = -0.4155(30)~~~~~~\gbr  =  0.1098(101)  \]
which may be compared with the SM predictions of:
\[ \gbl  = -0.4208~~~~~~\gbr  =  0.0774  \]
The value of $\gbl$ is quite consistent with the SM prediction
( for $m_t = 172$ GeV, $m_H = 149$ GeV ) 
( a 1.3$\%$, 1.8$\sigma$ deviation) whereas the discrepancy for $\gbr$ is much larger,
(a 42$\%$, 3.2$\sigma$ deviation).
Excluding the \tpol~ measurements gives the results:
\[ \gbl  = -0.4138(29)~~~~~~\gbr  =  0.1160(90)  \]
The deviations from the SM prediction are 1.7$\%$ and 2.4$\sigma$ for $\gbl$
and 50$\%$ and 4.3$\sigma$ for $\gbr$. 
 One may remark that the weak iso-spin of the SM 
affects only $\gbl$ not $\gbr$, so it is possible that the SM does correctly describe
$\gbl$ but that there is a new, anomalous, right handed coupling for the b quark.
\par The right- and left-handed effective couplings of the s,d quarks have recently been 
measured by the OPAL collaboration\cite{x34} with the results: 
\[ \overline{g}^L_{d,s} = -0.44_{-0.09}^{+0.13}
~~~~~~\overline{g}^R_{d,s} = 0.13_{-0.17}^{+0.15} \]
to be compared with the SM predictions -0.424 and 0.077 respectively. These
 measurements are
in good agreement with both the SM predictions and the measured b quark
couplings given above.
\par Limits can also be set on possible anomalous couplings of the
other `d-type' quarks, d,s by comparing the measured values of
$\langle A_{FB}^q \rangle$ and $\Gamma_{had}$ with the predictions of
 a model in which the 
d and s quarks are assumed to have the same effective coupling constants
as those measured for the b quarks. The prediction of this model for
$\langle A_{FB}^q \rangle $ is 0.1600(72), which is consistent with the
`measured' value (see Section 2 above and Table 8) of 0.1592(86) at the 
0.68$\sigma$ level. No useful constraint on possible anomalous couplings
of the d and s quarks, of a size similar to those observed for the b quark,
is therefore obtained using 
the measured value of $\langle A_{FB}^q \rangle $ with the present
experimental errors. A more favourable case is  $\Gamma_{had}$.
Using the world average value of $\alps$ of 0.118(5)~\cite{x9,x10} in Eqn.(2.17)
to calculate the QCD correction, and with $C_q^{QED} = 1.00040$, the predicted
value of $\Gamma_{had}$ in the model with a universal right-handed anomaly
for down-type quarks is 1.7249(46) GeV. This differs by 3.6$\sigma$ from
the LEP average measurement~\cite{x1} of 1.7436(25) GeV. Thus the model is 
essentially excluded by the measurement of $\Gamma_{had}$.
It is interesting to note that the precise measurement of $\Gamma_{had}$
currently 
gives a much more stringent constraint on possible anomalous couplings of the
d and s quarks than the direct measurement of their left- and right-handed
couplings cited above~\cite{x34}.
\par The values of the left- and right-handed couplings of the c quarks
derived from the measured values of $\abc$ and $\vbc$ given in the first
two rows of Table 7 are:  
\[ \gcl  = 0.3440(92)~~~~~~\gcr  =  -0.1600(70)  \]
in very good agreement with the SM predictions
for $m_t = 172$ GeV, $m_H = 149$ GeV of:
\[ \gcl  = 0.3465~~~~~~\gcr  =  -0.1545.  \]
The $\pm 2 \sigma$ limits for deviations of $\gcr$ from the SM prediction
extends from -0.174 to -0.146. Thus, at 95$\%$ CL, any anomalous 
right-handed couplings of the c quark lie between -9$\%$ and +15$\%$ of the
SM prediction.

\SECTION{\bf{Summary and Outlook}}
The analysis presented in this paper has been performed in 
such a way as to separate,
as far as possible, the actual results of experimental measurements,
expressed in terms of the effective weak coupling constants of the leptons,
c quarks and b quarks, from the comparison of these results with the SM.
 The quantities $\vbl$, $\abl$ and $\rbQ = \vbQ/\abQ$ (~$Q=$c,b~) were
 extracted assuming only lepton universality. A further theoretical assumption
 is needed to extract $\vbQ$ and $\abQ$ separately. The hypothesis of 
 non-b quark lepton universality was chosen. It was shown that
 the alternative of assuming
 $\alpha_s(M_Z)$ to be known\footnote{This was the procedure adopted
 in Ref.[7], where where heavy quark effective couplings consistent with
 those extracted in the present paper were found.}, would give essentially the same results
 for the b quark couplings, if the world average value of
 $\alpha_s(M_Z) = 0.118(5)$~\cite{x9,x10} is used. Thus the actual non-b quark 
 couplings are consistent with the universality hypothesis.
 \par The measured effective couplings of the leptons and c quarks are in
 good agreement with the SM predictions for a values of $m_t$ and $m_H$
well consistent with existing experimental limits. However, the b quark couplings deviate from the SM predictions by
 more than three standard deviations. Taking carefully into account all relevant
 error correlations, the probablity that all six effective couplings are consistent with
 lepton universality and the SM is found to be 0.9$\%$. This number is the 
 product of the CL for consistency with lepton universality,
 given by the $\chi^2$ of the different $A_l$ when compared 
 with their weighted average (8.4 $\%$) and the CL
 of the SM comparison using the average value from Table 9 (10.5 $\%$).
  The probability that the leptonic and
 b quark couplings are consistent with lepton universality and the SM
 is similarly calculated to be 0.18$\%$. The latter probability
  drops to only
  0.018$\%$ if the \tpol~ measurements are excluded in calculating the
  average value of $A_l$. 
   \par In Section 3 the possibility is discussed that systematic effects,
 beyond those taken into account in the present analyses and    
   correlated between different experiments, exist in the \tpol~ measurements
 \par In Section 4 the measured effective couplings are analysed in terms of
  the one-loop quantum correction parameters $\Delta \rho_f$ and 
  $\Delta \kappa_f$. These quantities are found to be in agreement
  with the SM predictions for the leptons and c quarks, but to be an order of
  magnitude larger than these predictions for the b quarks. An analysis in terms
  of the `epsilon' parameters~\cite{x30,x31,x32} $\epsilon_1$,  $\epsilon_2$,
  $\epsilon_3$, $\epsilon_b(\sbb)$, $\epsilon_b(\abb)$ and $\epsilon_b(\rbb)$
  shows reasonable agreement with the SM for the first four parameters, but
  $\simeq 4-6 \sigma$ deviations for the last two. The agreement
  found between
  experiment and the SM for $\epsilon_1$, $\epsilon_2$ and $\epsilon_3$ 
  demonstrates that there is no experimental evidence for contributions
  from natural extensions of the SM such as Supersymmetry, Technicolour,
  anomalous WW$\gamma$ or WWZ couplings, or new U(1) gauge bosons, which
   are all expected to produce deviations from the SM model
  predictions for one or more of these parameters~\cite{x31}.
 It may be remarked, however, that if the \tpol~ measurements are excluded
 the  $\epsilon_3$ parameter differs from the SM prediction by three
standard deviations, even for a Higgs boson mass ($m_H = 100$ GeV) close
to the current experimental lower limit~\cite{x28} of about 70 GeV.
 This additional potential problem for the SM merits further
investigation, but is beyond the scope of the present paper. 
  \par A simple picture emerges in terms of the right-handed and left-handed
  couplings of the b quark. While the latter agrees
  at the 2$\sigma$ level or better with the SM
  prediction, the former shows a 42-50$\%$ and 3.2-4.3$\sigma$ deviation.  
  The larger deviations occur when \tpol~ measurements are excluded.
   Thus the most significant deviation from the SM predictions of the effective couplings
  is found in the right-handed b quark weak
  coupling constant. 
  \par The large deviations from the SM found here for the b quark couplings
  confirm results presented in a previous paper~\cite{x7}. The paper contains,
  however, neither
  any discussion of the overall statistical significance of the deviations nor any
  physical interpretation of them. Other recent reviews~\cite{x33,x35,x36} did not 
  extract the heavy quark couplings from the measurements. They discussed rather
  the high precision purely leptonic data and the quantities $R_c$ and $R_b$ all
  of which agree well with SM predictions. Of the three parameters
  $\epsilon_b(\sbb)$, $\epsilon_b(\abb)$ and $\epsilon_b(\rbb)$ defined in
  Ref.[32], only $\epsilon_b(\sbb)$, derived from $R_b$, was considered in Ref.[33].
  As also only the ALEPH $R_b$ measurement~\cite{x37} (which is in almost perfect 
  agreement with the SM) was used, the anomalies in the b quark couplings were
  undetected. As shown in Tables 16 and 17 it is the other two $\epsilon_b$ parameters 
  that are most sensitive to the apparent deviations from the SM.
  \par The most important remaining question is whether the observed deviation
  in the right-handed b quark coupling is likely to be confirmed or excluded 
  by currently planned measurements of electroweak observables. The effect 
  observed in the effective couplings, interpreted as a statistical
  fluctuation, is $\simeq$ 3-4$\sigma$.
   A similar sized deviation for $R_b$ was reported in 1995 by the
  Electroweak Working Group~\cite{x38}. A year later, as a result of a 
  better systematic understanding of both the $R_b$ and the correlated
  $R_c$ measurements, as well as much improved statistical errors, the deviation
  of the average $R_b$ measurement was reduced to below two standard
  deviations~\cite{x1}. Can the presently observed anomaly in the b quark
  right-handed coupling be expected to meet the same fate as that of $R_b$?
   Arguments will now be given that this is unlikely, given the current
   status of the analysis of the LEP+SLD electroweak data. If there is 
   some large, as yet unknown, systematic effect in the LEP $A_{FB}^{0,b}$
   and/or the SLD $A_b$ measurements similar to those uncovered during
   1996 in the $R_b$ and $R_c$ measurements, the anomaly might disappear 
   (or become larger). This hypothetical question will not be further
   addressed here.
   \par The discussion concentrates on the observed (and 
   possible future) deviations from the SM of $\rbb$, which is extracted
    from measurements
   using only the weak theoretical assumption of lepton universality.
   As can be seen in Figs.1c,2c this quantity shows much larger deviations from the
   SM predictions than $\sbb$.
   The measurements of several of the relevant electroweak observables:
   $A_{FB}^{0,l}$, $A_{LR}$, $A_{FB}^{0,b}$ and $A_b$ have recently been 
   improved as compared to numbers quoted in Table 1~\cite{x39}.
   The updated value of $\rbb$ following the same analysis procedure as
   described in Section 2 above is 0.591(30). This is very consistent with 
   the value reported in Table 3. The discrepancy with the SM is slightly
   reduced from 3.34 to 3.26 standard deviations. The forseen improvements
   in the precision of measurements of electroweak observables have been
   discussed in detail in the Review of Renton~\cite{x7}. Only very modest
   improvements are to be expected from completing the analysis of the
   existing and final LEP1 data. Their impact on the $\rbb$ 
   measurement is expected to be essentially negligible. As shown in
   Sections 2 and 3 above this is not true, however of possible
   systematic effects, witness
   the sensitivity of the CLs in Table 9 
   and the $\Delta \rho$,  $\Delta \kappa$ and
   epsilon parameters in Tables14-17,to the inclusion or exclusion of the
   \tpol~ data. A better systematic understanding of the latter could 
   have a large effect on the LEP average value of $A_l$. Ultimately however
   the LEP+SLD average value of $A_l$ is expected to be dominated by the
   improved precision of the ongoing $A_{LR}$ measurement. According to Ref.[7]
   the errors on the existing SLD measurements of $A_{LR}$ and $A_b$~\cite{x39}
   should be reduced, by the end of the SLD experimental program, by
   41$\%$ and 45$\%$ respectively. The most significant improvement is expected to
   be in the accuracy of the LEP+SLD average value of $A_l$ due to the $A_{LR}$
   measurement. The error of the latest measurement: $A_l^{SLD} =$ 0.1547(32)~\cite{x39}
   is expected to be reduced to 0.0019. This is significantly smaller than the
   error, 0.0033, on the LEP average value of $A_l$ which will,
in the absence of new, presently unknown, systematic corrections,
  change little
   in the future. The improvement on the most recent SLD measurement of $A_b$,
   $A_b^{SLD} =$ 0.897(47)~\cite{x39}, where the error is expected to be reduced to
   0.0027, is also large. However, the LEP average value of $A_b$, derived from
   $A_{FB}^{0,b}$ is already more precise: $A_b^{LEP} =$ 0.861(21). 
   To study the possible impact of the new SLD measurements on the 
   expected future value of $\rbb$ the value of $A_l^{SLD}$ is allowed to vary
   by $\pm2 \sigma_o$ (where  $\sigma_0 = 0.0019$ is the final expected error)
   about the latest measured value. The weighted average of $A_l^{SLD}$ and the
   latest LEP measurement, is calculated (including, or not, the \tpol~ data)
   to estimate the likely range of the `final' LEP+SLD average value of $A_l$.
   For each $A_l$ value, $A_b$ is extracted from the LEP average $A_{FB}^{0,b}$
   using Eqn.(2.3) to yield `final' values of $A_b^{LEP}$. The weighted 
   average of $A_b^{LEP}$ and  $A_b^{SLD}$ is then made, assuming for the
   latter the latest value given above and the expected final error of
  0.0027. In the last step, for each `final' $A_b$ value $\rbb$ is 
  extracted using Eqn.(2.6) and compared with the SM prediction. The results of
  this exercise are presented in Table 18. This purely statistical study shows
  that the deviations in $\rbb$ are unlikely to drop below 3$\sigma$ , or to 
  become more significant than 5$\sigma$. Larger deviations are still
  seen when the \tpol~ data are excluded, but the effect is less than
  shown in Table 3 ($\simeq  0.5 \sigma$ instead of $\simeq  1 \sigma$) due to
  the much higher statistical weight expected from the final SLD $A_l$
  measurement. Unlike in the case of the $R_b$ anomaly in 1995, most of the
  foreseeable LEP+SLD data contributing to the measurement of $\rbb$ have 
  already been analysed. The predictions presented in Table 18 show that
  only small changes are to be expected in the statistical significance of the 
  observed anomaly in the b quark couplings.

  \par The most important message of this paper for future electroweak 
  analyses is that the step of extracting the important physical
  quantities from the experimental measurements should be clearly separated
  from the comparison of these quantities with theoretical predictions.
  This was done routinely in the past for the leptonic weak coupling
  constants~\cite{x1}.
  A similar procedure should be followed, in future, also for the heavy quark
  couplings. Only in this way can the physical origins of apparent
  deviations from theory be precisely located, and the CLs for 
  agreement of the measurements with the theory be easily and
  correctly calculated. The second point (exemplified here by the 
  discussion of the LEP \tpol~ measurements) is to check carefully
  the internal consistency of different experimental measurements of
  the same physical quantity before 
  they are averaged. This can give indications of hitherto
  unconsidered systematic effects. In any case, the CL for the internal
  consistency of different measurements used to calculate an average
  physical quantity should be multiplied by that given by
  the $\chi^2$ of the theory/experiment
  comparison based on the averaged quantities, in order to give a
  more meaningful
  overall probability for the theory/experiment comparison. 

\section{Acknowledgements}
I thank M.Consoli, W.Lohmann and Z.W\c{a}s for helpful discussions, and
the referee for pointing out an important numerical error in the QED 
corrections in the original version of the paper.
%\newline  
%\pagebreak

\pagebreak
\begin{table}
\begin{center}
\begin{tabular}{|c|c|c|c|} \hline
  Quantity  &  Measurement (Total Error) & SM & (Meas.-SM)/Error \\
\hline  
  LEP        &   &   &    \\  \cline{1-1}
  $A_{FB}^{0,e}$   & 0.0160(24) & 0.0159 & 0.04  \\
  $A_{FB}^{0,\mu}$   & 0.0162(13) & 0.0159 & 0.04  \\
  $A_{FB}^{0,\tau}$   & 0.0201(18) & 0.0159 & 2.3  \\ 
  $\Gamma_l$ (Mev)   & 83.91(11) & 83.96 & -0.45  \\ 
  \tpol     &  &  & \\
  $A_e$  &   0.1382(76) & 0.1458 & -1.0 \\
  $A_{\tau}$  &   0.1401(67) &  0.1458 & -0.9 \\
  c and b quarks  &  &  &  \\
   $A_{FB}^{0,c}$   & 0.0733(49) & 0.0730 &  0.1   \\ 
   $R_c$            & 0.1715(56)  & 0.1723 & -0.1 \\
   $A_{FB}^{0,b}$   & 0.0979(23)  & 0.1022 & -1.8  \\ 
   $R_b$            & 0.2179(12)  & 0.2158 &  1.8 \\ 
\hline 
  SLD   &  &  & \\ \cline{1-1}
  $A_e$  &   0.1543(37)  & 0.1458 & 2.3 \\
  $A_c$  &   0.625(84) & 0.667 & -0.5 \\
  $A_b$  &   0.863(49) & 0.935 & -1.4 \\
  $R_b$  &   0.2149(38) & 0.2158 & -0.2 \\              
\hline
\end{tabular}
\caption[]{ Average values of electroweak observables used in the analysis
~\cite{x1}. SM denotes the Standard Model prediction for $m_t =$ 172 GeV, 
$m_H =$ 149 GeV~\cite{x1}}
\end{center}
\end{table} 
\begin{table}
\begin{center}
\begin{tabular}{|c|c|c|c|c|} \hline
  $A_l$  & $A_c$  & $A_b$  & $R_c$ &  $R_b$  \\  
\hline
0.1501(24) & 0.645(39) & 0.869(22) & 0.1715(56) & 0.2176(11) \\
0.1533(27) & 0.634(38) & 0.853(22) & - & -  \\
\hline
\end{tabular}
\caption[]{ LEP+SLD averages. The \tpol~  measurements are excluded from
the averages quoted in the second row }      
\end{center}
\end{table}
\begin{table}
\begin{center}
\begin{tabular}{|c|c|c|c|} \hline
      & $\rbl$  & $\rbc$  &  $\rbb$   \\
 \hline      
Measurement &  0.07548(120) &  0.366(29)   & 0.582(32)  \\
     SM &  0.07332 &  0.383   & 0.689  \\
 (Meas.-SM)/Error&  -1.80  & -0.59 & -3.34  \\
 \hline
Meas. $\tau$ poln. out &  0.07711(140) &  0.357(29)   & 0.562(29)  \\
(Meas.-SM)/Error&  -2.71  & -0.90 & -4.38  \\   
\hline
\end{tabular}
\caption[]{ Measured values of $\rbf = \vbf/\abf$ compared to Standard Model 
predictions } 
\end{center}
\end{table}
\begin{table}
\begin{center}
\begin{tabular}{|c|c|c|c|} \hline
      & $\sbl$  & $\sbc$  &  $\sbb$   \\
\hline      
Measurement &  0.25244(33) &  0.2877(95)   & 0.3676(24)  \\
     SM &  0.25259 &  0.2880   & 0.3644  \\
 (Meas.-SM)/Error &  -0.45  & -0.03 & 1.33  \\
\hline
\end{tabular}
\caption[]{ Measured values of $\sbf = \abf^2(1-6 \mu_f)+\vbf^2$  compared 
to Standard Model predictions } 
\end{center}
\end{table}
\begin{table}
\begin{center}
\begin{tabular}{|c|c|c|c|} \hline
      & \multicolumn{3}{c|}{leptons}  \\ \cline{2-4}       
      & Meas. & SM & Dev($\sigma$)  \\
\hline  
$\abl$ & -0.50101(33) & -0.50124 &  0.67 \\ 
$\vbl$ & -0.03782(68) & -0.03675 &  -1.57 \\              
 \hline
$\abl$ & -0.50093(33) & - &  0.91  \\ 
$\vbl$ & -0.03863(77) & - &  -2.44 \\               
 \hline 
\end{tabular}
\caption[]{ Measured values of the effective electroweak coupling 
constants for the charged leptons. Dev($\sigma$)  = (Meas.-SM)/Error.
The values given in the 
last two rows exclude \tpol~  data from the averages }
\end{center}
\end{table}
\begin{table}
\begin{center}
\begin{tabular}{|c|c|c|c|} \hline
  $C_c^{QED}$ & $C_b^{QED}$  & $C_c^{QCD}$  & $C_b^{QCD}$   \\
\hline      
 1.00046 &  0.99975 &  1.0012   & 0.9953  \\
\hline
\end{tabular}
\caption[]{ QED and QCD correction factors for heavy quarks
assuming $\alpha_s(M_Z) = 0.12$ and $\alpha(M_Z)^{-1} = 128.9$.  } 
\end{center}
\end{table}
\begin{table}
\begin{center}
\begin{tabular}{|c|c|c|c|c|c|c|} \hline
      & \multicolumn{3}{c|}{c quark} 
      &  \multicolumn{3}{c|}{b quark} \\  \cline{2-7}    
      & Meas. & SM & Dev($\sigma$)
      & Meas. & SM & Dev($\sigma$)  \\
\hline  
$\abf$ & 0.504(10) & 0.501 & 0.30 
       & -0.5252(75) & -0.4981 & -3.61  \\ 
$\vbf$ & 0.184(15) & 0.192 & -0.53 
       & -0.3057(125) & -0.3434 & 3.18  \\              
 \hline
$\abf$ & 0.505(10) & - & 0.40 
       & -0.5298(70) & - & -4.53  \\ 
$\vbf$ & 0.180(15) & - & -0.80 
       & -0.2977(123) & - & 3.72  \\              
 \hline 
\end{tabular}
\caption[]{ Measured values of the effective electroweak coupling 
constants of c and b quarks. Dev($\sigma$) = (Meas.-SM)/Error.
 The values given in the 
last two rows exclude \tpol~  data from the averages }
\end{center}
\end{table}
\begin{table}
\begin{center}
\begin{tabular}{|c|c|c|c|c|} \hline
   &`Measured' &  SM Pred. & MI Pred.  & MI Pred. with SM b quark \\
\hline      
 $\langle A_{FB}^q \rangle$  & 0.1592(86) & 0.1641 & 0.1639(28) & 
 0.1692(28)  \\
\hline
(`Meas'.-Pred.)/Error &  & -0.57 & -0.52  & -1.1 \\
\hline
\end{tabular}
\caption[]{ Values of the mean quark charge asymmetry. `MI Pred.'
stands for Model Independent Prediction (see text). See also
the text for the definition of `measured'. }
\end{center}
\end{table} 
\begin{table}
\begin{center}
\begin{tabular}{|c|c|c|c|c|c|} \hline
  Observables & $\rbl$, $\sbl$ & $\rbl$, $\rbb$  &
   $\rbl$, $\sbl$, $\rbb$, $\sbb$ & $\rbl$, $\rbb$, $\rbc$ &
   $\rbl$, $\sbl$, $\rbb$, $\sbb$, $\rbc$, $\sbc$  \\
\hline  
 $dof$ & 2 & 3 & 5 & 5 & 8 \\
 $\chi^2$ & 3.44 & 10.6 & 13.2 & 10.9 & 13.2 \\
 CL ($\%$) & 17.9 & 1.4 & 2.2 & 5.3 & 10.5 \\
\hline 
$\chi^2$ & 7.55 & 17.2 & 19.4 & 17.4 & 19.6 \\
 CL ($\%$) & 2.3 & 0.064 & 0.16 & 0.38 & 1.2 \\     
\hline
\end{tabular}
\caption[]{ $\chi^2$ and confidence levels for agreement with the SM
( $m_t =$ 172 GeV, $m_H =$ 149 GeV)
of different sets of electroweak observables sensitive to the
effective couplings, assuming perfect statistical consistency
 of the LEP+SLD averages in Table 2. 
The values given in the last two rows do not
use \tpol~measurements in the $A_l$ average.
See the text for the explanation of
 the number of degrees of freedom ($dof$) in each case.}
\end{center}
\end{table}
\begin{table}
\begin{center}
\begin{tabular}{|c|c|c|c|} \hline
      & $\tau-\mu$ & $\tau-e$ & $\mu-e$  \\
 \hline
CL($\%$) STT & 2.7 & 6.0 & 68  \\
 \hline
CL($\%$) EET & 8.0 & 17 & 94 \\
 \hline 
\end{tabular}
\caption[]{ Confidence levels for the consistency of LEP measurements of
 $A_{FB}^{e,0}$, $A_{FB}^{\mu,0}$ and $A_{FB}^{\tau,0}$. STT: Student's t test, EET:
 CL calculated from Estimated Experimental errors. }
\end{center}
\end{table}   
\begin{table}
\begin{center}
\begin{tabular}{|c|c|c|c|c|c|} \hline
    & ALEPH & DELPHI & L3  & OPAL & WA \\
\hline 
\hline
$e \nu \nu$ & 0.200(51,31) & 0.179(52,67) & 0.168(39,15) &
 0.161(33,29) & 0.173(26) [16] \\
$\mu \nu \nu$ & 0.124(41,21) & 0.097(38,22) & 0.111(45,16) &
 0.138(33,22) & 0.119(22) [11] \\ 
$\pi \nu$ & 0.142(20,11) & 0.158(33,50) & 0.135(21,17) &
 0.117(14,12) & 0.130(12) [9] \\
$\rho \nu$ & 0.108(19,18) & 0.199(46,39) & 0.168(17,10) &
 0.116(13,11) & 0.135(12) [22] \\
$a_1 \nu$ & 0.135(35,20) & 0.103(50,38) &  -  &
 0.151(37,31) & 0.134(28) [14] \\
\hline 
 WA  & 0.132(15) [24] & 0.137(26) [27] & 0.154(14) [14]   &
 0.123(11) [9] & 0.1343(73) \\
\hline        
\end{tabular}
\caption[]{ LEP measurements of $A_{\tau} = -\langle P_{\tau} \rangle$
The first error shown is statistical, the second systematic. For the
Weighted Averages (WA) the weighted total error is given, where statistical
and systematic errors are added in quadrature. Errors calculated from sample
variances are shown in square brackets. }
\end{center}
\end{table} 
\begin{table}
\begin{center}
\begin{tabular}{|c|c|c|c|c|c|c|c|c|c|c|} \hline
      & $e-\mu$ & $e-\pi$ & $e-\rho$ & $e-a_1$ & $\mu-\pi$
& $\mu-\rho$ & $\mu-a_1$ & $\pi-\rho$ & $\pi-a_1$ & $\rho-a_1$  \\
 \hline
CL($\%$) STT & 0.3 & 1.5 & 36 & 2.5 & 15 & 35 & 9 & 68 & 45 & 95 \\
 \hline
CL($\%$) EET & 11 & 13 & 19 & 31 & 67 & 51 & 66 & 74 & 79 & 97 \\
 \hline 
\end{tabular}
\caption[]{ Confidence levels for the consistency of LEP
 $\langle P_{\tau} \rangle$
measurements using different $\tau$ decay channels. STT and EET are defined in Table 10.}
\end{center}
\end{table}
\begin{table}
\begin{center}
\begin{tabular}{|c|c|c|c|c|c|c|} \hline
      & A-D & A-L & A-O & D-L &  D-O & L-O \\
 \hline
CL($\%$) STT & 85 & 31 & 15 & 31 & 31 & 64 \\
 \hline
CL($\%$) EET & 87 & 29 & 63 & 57 & 63 & 9 \\
 \hline 
\end{tabular}
\caption[]{ Confidence levels for the consistency of LEP
 $\langle P_{\tau} \rangle$
measurements by different experiments. STT and EET are defined in Table 10. }
\end{center}
\end{table}
\begin{table}
\begin{center}
\begin{tabular}{|c|c|c|c|c|c|c|} \hline
      & \multicolumn{2}{c|}{leptons} 
      & \multicolumn{2}{c|}{c quark}
      & \multicolumn{2}{c|}{b quark} \\  \cline{2-7}    
      & $\Delta \rho_l$ & $\Delta \kappa_l$ &  $\Delta \rho_c$
      & $\Delta \kappa_c$ & $\Delta \rho_b$ & $\Delta \kappa_b$  \\
\hline  
Expt. & 0.00404(133) & 0.03445(134) & 0.016(41) 
       & 0.064(49) & 0.101(32) & 0.403(107)  \\
\hline
SM $m_t =$ 172 GeV&   &   &   &   &   & \\         
 $m_H =$ 149 GeV & 0.00497 & 0.03686 & 0.005 
       & 0.037 & -0.007 & 0.0436  \\
 Dev($\sigma$) & -0.7 & -1.8 & 0.27 & 0.55 & 3.38 & 3.36 \\                     
 \hline
SM $m_t =$ 180 GeV&   &   &   &   &   & \\     
   $m_H =$ 100 GeV & 0.00563 & 0.03472 & 0.006 
       & 0.034 & -0.008 & 0.0412  \\
 Dev($\sigma$) & -1.2 & -0.02 & 0.24 & 0.61 & 3.40 & 3.38 \\                     
 \hline 
\end{tabular}
\caption[]{ Measured values of the quantum correction parameters
$\Delta \rho_f$ and $\Delta \kappa_f$ compared to SM predictions.
 Dev($\sigma$) = (Expt.-SM)/Error. }
\end{center}
\end{table}
 
\begin{table}
\begin{center}
\begin{tabular}{|c|c|c|c|c|c|c|} \hline
      & \multicolumn{2}{c|}{leptons} 
      & \multicolumn{2}{c|}{c quark}
      & \multicolumn{2}{c|}{b quark} \\  \cline{2-7}    
      & $\Delta \rho_l$ & $\Delta \kappa_l$ &  $\Delta \rho_c$
      & $\Delta \kappa_c$ & $\Delta \rho_b$ & $\Delta \kappa_b$  \\
\hline  
Expt. & 0.00372(133) & 0.03260(157) & 0.020(41) 
       & 0.079(49) & 0.119(30) & 0.470(97)  \\
\hline 
 SM $m_t =$ 172 GeV&   &   &   &   &   & \\              
 $m_H =$ 149 GeV & 0.00497 & 0.03686 & 0.005 
       & 0.037 & -0.008 & 0.0439  \\
 Dev($\sigma$) & -0.94 & -2.7 & 0.37 & 0.86 & 4.23 & 4.39 \\                     
 \hline
 SM $m_t =$ 180 GeV&   &   &   &   &   & \\     
   $m_H =$ 100 GeV & 0.00563 & 0.03472 & 0.006 
       & 0.034 & -0.008 & 0.0412  \\
 Dev($\sigma$) & -1.4 & -1.4 & 0.34 & 0.92 & 4.23 & 4.42 \\   

 \hline 
\end{tabular}
\caption[]{ Measured values of the quantum correction parameters
$\Delta \rho_f$ and $\Delta \kappa_f$ compared to SM predictions.
\tpol~ measurements are excluded from the $A_l$ average.
 Dev($\sigma$) = (Expt.-SM)/Error. }
\end{center}
\end{table}

\begin{table}
\begin{center}
\begin{tabular}{|c|c|c|c|c|c|c|} \hline
  & $\epsilon_1$ & $\epsilon_2$ & $\epsilon_3$ & $\epsilon_b(\sbb)$
  & $\epsilon_b(\abb)$ & $\epsilon_b(\rbb)$ \\
\hline  
Expt. & 0.00404(133) & -0.0073(8) & 0.0031(8) 
       & -0.0017(18) & 0.048(15) & -0.263(57)  \\
\hline
SM $m_t =$ 172 GeV&   &   &   &   &   & \\         
   $m_H =$ 149 GeV & 0.00497 & -0.0076 & 0.0051 
       & -0.0045 & -0.0060 & -0.0068  \\
 Dev($\sigma$) & -0.7 & 0.38 & -2.5 & 1.6 & 3.6 & -4.5 \\                     
 \hline
SM $m_t =$ 180 GeV&   &   &   &   &   & \\  
   $m_H =$ 100 GeV & 0.00563 & -0.0062 & 0.0045 
       & -0.0046 & -0.0068 & -0.0055  \\
 Dev($\sigma$) & -1.2 & -1.4 & -1.8 & 1.6 & 3.7 & -4.5 \\                     
 \hline 
\end{tabular}
\caption[]{ Measured values of the epsilon parameters of 
Refs.[16-18] compared to SM predictions.
 Dev($\sigma$) = (Expt.-SM)/Error. }
\end{center}
\end{table}

\begin{table}
\begin{center}
\begin{tabular}{|c|c|c|c|c|c|c|} \hline
  & $\epsilon_1$ & $\epsilon_2$ & $\epsilon_3$ & $\epsilon_b(\sbb)$
  & $\epsilon_b(\abb)$ & $\epsilon_b(\rbb)$ \\
\hline  
Expt. & 0.00372(133) & -0.0068(8) & 0.0021(8) 
       & -0.0018(18) & 0.058(14) & -0.298(47)  \\
\hline
SM $m_t =$ 172 GeV&   &   &   &   &   & \\         
   $m_H =$ 149 GeV & 0.00497 & -0.0076 & 0.0051 
       & -0.0045 & -0.0060 & -0.0068  \\
 Dev($\sigma$) & -0.94 & 1.0 & -3.8 & 1.5 & 4.6 & -6.2 \\                     
 \hline
SM $m_t =$ 180 GeV&   &   &   &   &   & \\  
   $m_H =$ 100 GeV & 0.00563 & -0.0062 & 0.0045 
       & -0.0046 & -0.0068 & -0.0055  \\
 Dev($\sigma$) & -1.4 & -0.75 & -3.0 & 1.6 & 4.6 & -6.2 \\                     
           
 \hline
\end{tabular}
\caption[]{ Measured values of the epsilon parameters of 
Refs.[30-31] compared to SM predictions.
\tpol~ measurements are excluded from the $A_l$ average.
 Dev($\sigma$) = (Expt.-SM)/Error. }   
\end{center}
\end{table} 
\begin{table}
\begin{center}
\begin{tabular}{|c|c|c|c|c|c|c|} \hline
      & \multicolumn{3}{c|}{`final values' of $A_l$} 
      & \multicolumn{3}{c|}{ `final values' of $\rbb$ } \\ 
\hline          
$A_l^{SLD}\pm x$,~$x$ =  & -2$\sigma_0$  &  0  & +2$\sigma_0$ &
-2$\sigma_0$  &  0  & +2$\sigma_0$  \\
\hline  
LEP+SLD & 0.1501(16) & 0.1525(16) & 0.1553(16) & 0.604(26) & 
0.592(25) & 0.579(23)    \\ 
Full average &    &    &    & [-3.3] & [-3.9] & [-4.8] \\
\hline
$\tau$ poln.   & 0.1513(17) & 0.1545(17) & 0.1577(17) & 0.598(26) & 
0.583(24) & 0.568(23)    \\ 
Excluded  &    &    &    & [-3.5] & [-4.4] & [-5.3] \\
\hline        
\end{tabular}
\caption[]{ A scenario for the `final' values and errors of
$A_l$ and $\rbb$ at the end of the LEP+SLD experimental programme.
$A_l^{SLD}= 0.1547(19)$ and $A_b^{SLD}= 0.897(27)$ are assumed, and
$A_l^{SLD}$ varies by -2$\sigma_0$ to +2$\sigma_0$ from the central
value, where $\sigma_0$ is its error. For $\rbb$, the number in square
brackets denotes: (Value-SM)/Error, where $\rbb^{SM} = 0.689$. }
\end{center}
\end{table} 
\end{document}